\documentclass[a4paper,fleqn,usenatbib]{mnras}

\usepackage[T1]{fontenc}
\usepackage{ae,aecompl}

\usepackage{graphicx}	
\usepackage[caption = false]{subfig}
\usepackage{amsmath}	
\usepackage{amssymb}	
\usepackage{multirow}
\usepackage{pdflscape}
\usepackage{deluxetable}

\title[HI angular momentum mass relation]{The HI angular momentum-mass relation}

\author[Kurapati et al.]{
Sushma Kurapati$^{1,2}$\thanks{E-mail: sushma@ncra.tifr.res.in},
Jayaram N. Chengalur $^{1}$ , Marc A.W. Verheijen $^{3,1}$ \\
$^{1}$ National Centre for Radio Astrophysics, Tata Institute of Fundamental Research, PO Box 3, Pune 411007, India \\
$^{2}$ The Inter-University Centre for Astronomy and Astrophysics, Post Bag 4, Ganeshkhind, Pune,  411007, India \\
$^{3}$ Kapteyn Astronomical Institute, University of Groningen, P.O. Box 800,
9700 AV Groningen, Netherlands }

\date{Accepted XXX. Received YYY; in original form ZZZ}

\pubyear{2018}

\begin{document}
\label{firstpage}
\pagerange{\pageref{firstpage}--\pageref{lastpage}}
\maketitle

\newcommand{\HI}{\rm H{\sc i }}
\newcommand{\MB}{\ensuremath{\rm M_B}}
\newcommand{\mb}{\ensuremath{\rm M_{bar}}}
\newcommand{\mhi}{\ensuremath{\rm M_{HI}}}
\newcommand{\mg}{\ensuremath{\rm M_{g}}}
\newcommand{\ms}{\ensuremath{\rm M_{s}}}
\newcommand{\msun}{\ensuremath{\rm M_{\odot}}}
\newcommand{\fatm}{\ensuremath{\rm f_{atm}}}
\newcommand{\jb}{\ensuremath{\rm j_{bar}}}
\newcommand{\js}{\ensuremath{\rm j_s}}
\newcommand{\jg}{\ensuremath{\rm j_{g}}}
\newcommand{\kms}{\ensuremath{\rm km \ s^{-1}}}
\newcommand{\kpc}{\ensuremath{\rm kpc}}
\newcommand{\LCDM}{\ensuremath{\Lambda{\rm CDM}}}
\newcommand{\aips}{{\sc aips }}
\newcommand{\gipsy}{{\sc gipsy }}
\newcommand{\fat}{{\sc fat }}
\begin{abstract}

We study the relationship between the \HI\ specific angular momentum (\jg) and the \HI\ mass (\mg) for a sample of galaxies with well measured \HI\ rotation curves. We find that the relation is well described by an unbroken power law \jg $\propto$ \mg$^{\alpha}$ over the entire mass range (10$^{7}$ -- 10$^{10.5}$ M$_{\odot}$), with $\alpha = 0.89 \pm 0.05$ (scatter 0.18 dex). This is in reasonable agreement with models which assume that evolutionary processes maintain \HI\ disks in a marginally stable state. The slope we observe is also significantly different from both the $j \propto M^{2/3}$ relation expected for dark matter haloes from tidal torquing models and the observed slope of the specific angular momentum-mass relation for the stellar component of disk galaxies. Our sample includes two  H{\sc i}--bearing ultra diffuse galaxies, and we find that their angular momentum follows the same relation as other galaxies. The only discrepant galaxies in our sample are early--type  galaxies with large rotating \HI\ disks which are found to have significantly higher angular momentum than expected from the power law relation.  The \HI\ disks of all these early--type  galaxies are misaligned or counter-rotating with respect to the stellar disks, consistent with the gas being recently accreted. We speculate that late stage wet mergers, as well as cold flows play a dominant role in determining the kinematics of the baryonic component of galaxies as suggested by recent numerical simulations.

\end{abstract}

\begin{keywords}
dwarf-galaxies: fundamental parameters--galaxies: kinematics and dynamics

\end{keywords}


\section{Introduction}

Studies of the correlations between global parameters of galaxies (e.g.
rotation velocity vs.  luminosity, or star formation rate
vs. stellar mass) provide critical observational tests of galaxy formation
and evolution models. One specific parameter which is receiving increasing 
attention is the angular momentum. This is  in part because it is only
now that, thanks to large IFU surveys as well as  large H{\sc i} surveys, measurements 
of angular momentum are becoming  available for a sizeable number of 
galaxies \citep[e.g.][]{walter08, hunter12, sanchez12, bryant15, Bundy15, scott16}. 
Further, our theoretical understanding of how galaxies acquire angular momentum is
also rapidly evolving as numerical models become more sophisticated and better
able to capture the details of the  physics of the baryonic component of galaxies.

The traditional understanding was that galaxies acquire their angular momentum
via tidal torquing of collapsing proto-halos by neighbouring mass concentrations.
In such models the bulk of the angular momentum gain happens around the turn around 
time of the proto-halo and the angular momentum remains roughly constant after that epoch.
The angular momentum of the final collapsed halo is generally quantified by  its `spin parameter' 
$\lambda_h = J_h|E_h|^{1/2}/(GM_h^{5/2})$ (where $J_h$ is the total angular momentum of 
the halo, $E_h$ the total energy, and $M_h$ the total mass). Both analytical calculations
(\citet{peebles69}), as well as numerical simulations (e.g. \citet{bett07,rodriguez16}) find 
that the spin parameter is roughly independent of the total mass of the halo. This 
leads to a scaling relation between the halo specific angular momentum
$j_h$ ($=J_h/M_h$) and the halo mass, viz. $j_h \propto M_h^{2/3}$. If one
assumes that the baryonic matter and dark matter have similar distributions
at early times and that galaxies form by cooling and angular momentum conserving
collapse of the baryonic component (\cite{fall83,mo98}) then a similar scaling
relation would be expected for the baryonic component of galaxies, viz.
\jb\ $\propto$ \mb$^{2/3}$. It has been found that such a scaling relation is indeed obtained for the stellar component of galaxies, although  individual galaxies can deviate significantly from the relation and the constant of proportionality varies between early and late type galaxies, as well as with bulge fraction\citep{fall83, romanowsky12, fall18} 

This traditional understanding has been challenged by several numerical
simulations  \citep[e.g.][]{desmond17, jiang19} of galaxy formation that include the effects of mergers and coldflows. These simulations show
that the collapsed baryonic component of galaxies does not arise solely from
the cooling and contraction of the initial hot baryonic component of the proto-halo. Both mergers as well as `cold' flows of baryonic material along filaments
that penetrate deep into  the halo \citep[e.g.][]{keres05,keres09, brooks09} could result in a significant change in the final baryonic mass as well as the final angular momentum. The effect on the angular momentum depends on whether the 
merger was `wet' (i.e. with a gaseous progenitor) or dry (i.e. with a gas poor
progenitor) as well as the  relative orientation of the spin of the incoming material vis a vis that of the  galaxy \citep[see e.g.][]{naab14, choi17, lagos18}. Additionally, the final baryonic mass as well as the angular momentum are affected by the energy input from  stellar winds and supernovae which could cause an outflow of material from the galaxy.  The strength of the  outflows  depends on whether the supernovae explosions  are clustered or not. Since  clustering of supernovae typically happens only near the  centres of galaxies (i.e. in regions where the material has low  specific angular momentum,  (sAM)) outflows lead to a preferential loss of low specific angular momentum material. This results  in an increase in the specific angular momentum  of the remaining material in the galaxy \citep{dutton09, brook11, sharma12}. As a consequence of these processes the angular momentum of the galaxy is not determined during its formation phase, but rather is affected by the entire evolutionary path of the galaxy.  Both the morphology as well as the angular momentum appear to be driven more by evolutionary processes and a galaxy's merger history than the original spin parameter of the dark matter halo  \citep[see e.g.][]{stewart17, zjupa17, jiang19,renzini20}. For example, simulations show that  accretion  of low sAM or counter rotating gas  leads to a central compact galaxy  ("blue nugget"), with an outer spheroidal stellar distribution, and  possibly an outer counter-rotating disk, while accretion  of co-rotating material typically leads to a disk type galaxy  with gas density  stable against rapid gravitational collapse  \citep[see e.g.][]{zolotov15,jiang19,kretschmer20}.

Recent observational studies have also begun to indicate that the relationship between the specific angular momentum and the mass is more nuanced than a simple $\jb \propto \mb^{2/3}$ scaling.  For the stellar \js -- \ms\ relation,  \citet{posti18, posti19} find a unbroken power-law \js $\propto$ \ms$^{\alpha}$ with $\alpha$ $\sim 0.55 \pm 0.02$ for all the galaxies in the stellar mass range of  10$^{7}$--10$^{11.5}$~M$_{\odot}$. On the other hand for the gas \jg -- \mg\ relation \citet{mancera21} have recently reported a slope of $ 1.01 \pm 0.04$.  Earlier studies have also found that gas rich dwarf galaxies appear to have an excess baryonic specific angular momentum compared to that expected from the scaling relation for bulgeless spirals given by \citet{obreschkow14}
{ \citep[e.g.][]{butler17, chowdhury17, kurapati18}.
\citet{kurapati18} also find that the elevation in specific angular momentum occurs for the galaxies with masses lower than $\sim 10^{9}$ M$_{\odot}$.  The scaling relation hence appears to depend both on the details of the galaxy morphology, as well as which component (stars, gas) of the galaxy is under consideration.

Here we study the relation between the baryonic specific angular momentum and mass, as well as that between the \HI specific angular momentum (\jg\ ) and \HI mass (\mg\ ) for a diverse sample of gas rich galaxies including  H{\sc i}-bearing  ultra-diffuse galaxies (HUDs), normal dwarf galaxies, spiral galaxies, and early--type  galaxies. We discuss the results in terms of the range of simulations discussed above. This paper is organized as follows. In \S \ref{obs}, we describe the sample and observations. In \S~\ref{results}, we determine the gas angular momentum and gas mass relation. We present a discussion in \S~\ref{sec:discussion}  and summarize the main results in \S~\ref{summary}. 

\section{Determination of specific angular momentum}
\subsection{Sample }
\label{obs}

Our sample consists of a wide variety of galaxies including dwarf galaxies, H{\sc i} bearing ultra diffuse galaxies, spiral galaxies, as well as early--type  galaxies. For some galaxies we present fresh measurements. We supplement this sample with measurements that are available in the literature. We discuss first the galaxies for which we present fresh measurements.

The H{\sc i} bearing ultra-diffuse galaxies (HUDs) were selected from the `more restrictive' sample of \citet{leisman17}, with half light radii r$_{\mathrm{g},\mathrm{eff}} > 1.5$ kpc, ${\mu }_{g,0} > 24$
mag~arcsec$^{-2}$, and g-band Magnitude  >  -16.8 mag, i.e. more extreme stellar sizes as compared to their stellar magnitudes. We used new Giant Metrewave Radio Telescope (GMRT) observations to obtain H{\sc i} morphologies and kinematics of 2 HUDs as part of a pilot study for a larger survey. We derive the rotation curves and H{\sc i} surface brightness profiles for these galaxies using Fully Automated TiRiFiC \citep[FAT;][]{kamphuis15} by fitting a tilted ring model to the data cube. The derived rotation curves and surface brightness profiles for ultra--diffuse galaxies are shown in Figures \ref{fig:rotcur1} and \ref{fig:sbr1} respectively.

Spiral galaxies are taken from the samples of \citet{verheijen01a}. \citet{verheijen01a}  presented results from a synthesis imaging survey of 43 galaxies in the nearby Ursa Major region using the Westerbork Synthesis Radio Telescope (WSRT). We require a sample of galaxies where rotation curves can be reliably derived. Hence, we consider only the galaxies that are well resolved across the major axis and exclude the galaxies that are in interacting pairs or that have disturbed velocity fields. 16 of 43 galaxies remain after the imposition of these criteria. We use K' band luminosity profiles (as  measured in \citet{verheijen01b}) to calculate the stellar mass profiles since the  uncertainties in the derivation of the mass to light ratio are much less in the  mid-infrared compared to the optical. We assume a constant K' band mass to light ratio of 0.24 to calculate the stellar surface density profiles \citep{ponomareva18}.  We use the rotation curves and the radial H{\sc i} surface density profiles from \citet{verheijen01b} to calculate the baryonic and gas specific angular momentum.

The early--type  galaxies were selected from the ATLAS$^{\rm 3D}$
\HI survey of a volume-limited, complete sample of 166 nearby early-type
galaxies (ETGs) brighter than M$_{\rm K}$ = -21.5 \citep{serra12}. In
constructing the sub-sample we first restrict the selection to 
galaxies with morphologies later than T > -4.0. \citet{serra12}
divided the galaxies with detected \HI into four morphological classes, viz. (i)~large 
discs - most of the \HI rotates regularly and is distributed in a disc or ring larger
than the stellar body of the galaxy, (ii)~small discs - most of the \HI rotates regularly 
and is distributed in a small disc confined within the stellar body of the galaxy,
(iii)~unsettled  - most of the \HI exhibits an unsettled morphology (e.g. tails or stream 
of gas) and kinematics, and (iv)~clouds - the \HI is found in small, scattered clouds around the galaxy. In order to calculate the angular momentum accurately, we require a sample where reliable rotation curves can be derived. Hence, we have considered only the galaxies from the first class, i.e. those with large discs that have regular velocity fields. We derived the rotation curve for these galaxies using FAT or $^{\rm 3D}$Barolo \citep{diteodoro15} on the final ATLAS$^{\rm 3D}$ data products, i.e \HI cubes with a typical angular resolution of $\sim$ 35 arcsecs.  We chose the galaxies that are well resolved across its major axis ($>$ 6 beams). Our final sample consists of those  galaxies for which  reliable rotation curves could be obtained by FAT or BAROLO.  BAROLO was used to derive the rotation curve for only one of the galaxies (NGC 3941), since the FAT derived rotation curve was unreliable.Two of the early--type galaxies (NGC3626 and NGC3941) have companions, which were excluded for the measurements of mass and specific angular momentum. The derived rotation curves and surface brightness profiles for the final sample of early--type galaxies are shown in Figures \ref{fig:rotcur2} and \ref{fig:sbr2} respectively. In Fig. \ref{fig:mom1} - \ref{fig:mom10},  we show (i) the integrated H{\sc i} intensity map contours overlaid on the optical image, (ii) Position-Velocity (PV) diagram taken along the major axis of the galaxy with the rotation curves overlaid on them. The dashed lines indicate the systemic velocity and kinematic center. The overplotted  violet triangles represent the rotation curve derived by 'FAT' (iii) the intensity weighted first moment of the  galaxy,  and  (iv)  velocity  field  of  the  best  fitting FAT model for all the HUDs and early-type galaxies. We find that all the rotation curves broadly follow the PV diagram. We note that PV diagrams are not fully representative of the rotation curve as they may reflect the effects of non-circular motion while FAT and other algorithms also take pixels off the kinematic major axis into account. 

  As mentioned above, in our analysis we also use the published data from various earlier studies; the details are given in the relevant sections.

\subsection{Measuring mass and specific angular momentum}
The H{\sc i} mass was calculated using the standard formula,  M$_{HI} $= 2.36$\times$ 10$^{5}$ $D^{2} \int$S $dv~ M_{\odot}$, where $D$ is the distance in Mpc, $S$ is the flux density in Jy and $dv$ is in km s$^{-1}$ \citep{roberts62}. 
The distances for early-type galaxies are taken from \citet{cappellari11}, which were obtained with the  surface  brightness  fluctuation (SBF) method. These distance estimates have an error of $\sim$ 7 -- 10 $\%$. 
The distance estimates for the UDGs are taken from \citet{leisman17}, which were calculated using the ALFALFA flow model, and the distance uncertainties due to proper motions are $\lesssim$ 15 $\%$. The distances for the Ursa Major spiral galaxies are taken from \citet{sorce14}, which were derived by using the Tully Fisher relation as an empirical distance estimator which provides a distance accuracy of $\sim$ 3--10 $\%$. We assume a conservative error of 15 $\%$ on distance measurements for all the galaxies. The quoted uncertainties on gas mass are based on the errors on the distance measurements as well as the error on flux densities. The error on the flux density is dominated by the calibration uncertainties, which are typically $\sim$ 10\%.
 We do not consider the mass to light ratio uncertainties while calculating the uncertainties of baryonic masses for the ultra-diffuse galaxies as they are dominated by the gas mass. We calculate the uncertainties of the baryonic masses for the UMa spirals (following \citep{lelli16}) by
\begin{equation}
\delta_{M_{\rm bar}} = \sqrt{\delta^{2}_{M_{\rm g}} + \bigg(\gamma_{\ast}\delta_{L_{\ast}}\bigg)^{2} + \bigg(L\delta_{\gamma_{\ast}}\bigg)^{2} + \bigg( 2M_{\rm bar} \frac{\delta_{D}}{D} \bigg)^{2}}
\end{equation}

where $\delta_{\rm g}$ and $\delta_{L}$, are the  errors on M$_{\rm g}$ and L due to uncertainties on total fluxes and $\delta_{\gamma}$ is the uncertainty in mass to light ratio. We assume $\delta_{\gamma_{\ast}}$ $\sim$ 0.1 dex for K'-band data \citep{ponomareva18}. 
The specific angular momentum j is computed by numerically evaluating the integral,
\begin{equation}
j \equiv \frac{J}{M} = \frac{\int_{0}^{R}dr 2\pi r^{2} \Sigma(r) v(r) cos[\delta i(r)]}{\int_{0}^{R}dr 2\pi r \Sigma(r)}
\end{equation}
where, $\Sigma(r)$ is the azimuthally averaged surface density, v(r) is the circular velocity at radius r, $\delta$i(r) is the difference in inclination with respect to the central disc inclination and R is the radius of the gas disk. This integral is evaluated separately for the stellar, the gaseous, and the baryonic components of the galaxy. The procedure used for the estimation of specific angular momentum is similar for the galaxies from this work and for the galaxies that are taken from the literature.
In this work, only the stellar, H{\sc i}, and Helium contributions to the baryonic mass are considered. Hence, the total baryonic surface density is $\Sigma_{\rm b}$ = $\Sigma_{\ast}$ + 1.35 $\Sigma_{HI}$ , where the 1.35 factor accounts for the He contribution. No correction was made for molecular hydrogen as well as ionized gas. The optical deprojected radial surface brightness profiles were derived by fitting elliptical annuli to the optical images by using the ELLINT task in the GIPSY package. The specific angular momentum of gas, stars and baryons was computed by numerically evaluating the integral in Equation (1). For all quantities, the integral converges within the error bars. The uncertainties on the specific angular  momentum  of the various  components  are  based  on  the errors on distance measurements, on rotation velocities, surface brightness values, and inclination values. The uncertainty in j$_{k}$ ($k$ being stars or gas) is given by:

\begin{equation}
{\footnotesize
\frac{\delta_{j_{k}}}{j_{k}} =  \sqrt{\big(\frac{\delta_{D}}{D}\big)^{2}+ \big(\frac{\delta_{i}}{tan(i)}\big)^{2} + \sum_{i}^{N}\big(\frac{\delta_{v_{i}}}{v_{i}} . \frac{J_{i}}{J}\big)^{2} + \sum_{i}^{N}\big(\frac{\delta_{S_{i}}}{S_{i}} .  \frac{J_{i}}{J}\big)^{2} }
}
\end{equation}

where $i$ is the inclination  ($\delta_{i}$ is its uncertainty). v$_{i}$ is the rotation velocity at each point on rotation curve ($\delta_{v_{i}}$ is its uncertainty), S$_{i}$ is the surface brightness value at each point on surface brightness profile ($\delta_{S_{i}}$ is its uncertainty), J$_{i}$ is the evaluated angular momentum in each ring and J is the total angular momentum of galaxy. The errors on specific angular momentum are often dominated by the errors on distance measurement.

 In Tables \ref{Table1} and \ref{Table2}, we present the measurements of stellar, gas, and baryonic mass and specific angular momentum of UMa spiral galaxies and ultra-diffuse galaxies respectively: The columns are as follows. Column (1) name of the galaxy; (2) stellar mass ; (3) gas mass; (4) total baryonic mass; (5) stellar specific angular momentum; (6) gas specific angular momentum and (7) baryonic specific angular momentum. In Table \ref{Table3} we show the properties and the measurements of the gas mass and gas specific angular momentum of the early--type galaxies. The columns are as follows. Column (1) name of the galaxy; (2) morphology ; (3) distance in Mpc (4) gas mass;  (5) gas specific angular momentum and (6) Notes on galaxies.

\subsection{Convergence criteria} 

We show the cumulative specific angular momentum profiles for 2 HUDs (both \jb and \jg profiles), 16 UMa spirals (both \jb and \jg profiles), and 8 early-type galaxies (only \jg profiles) in Figures \ref{fig:prof1}, \ref{fig:prof2}, and \ref{fig:prof3} respectively. We normalize the specific angular momentum and radius axes to allow the comparison between the profiles.
  We also make a quantitative assessment  of the convergence of these profiles. We follow \citet{posti18} to determine whether the specific angular momentum of a galaxy has converged or not. We select galaxies satisfying the following criteria:
\begin{equation}
\frac{j(<R_{N})-j(R_{N-1})}{j(R_{N})} < 0.1 \ \  \& \& \ \
\frac{d \ log(j(<R))}{d \ log(R)} \bigg|_{R_{N}} < 0.5
\end{equation}

i.e. that the last two points of the j profile differ by less than 10 $\%$ and that the logarithmic slope of the j profile in the outer-most point is less than 1/2. We find that all the galaxies except two early-type galaxies (NGC3626 and NGC6798) follow the convergence criterion.

We also fit a function $\mathcal{P}$ = k. exp(-a/x), (which is zero at x$\xrightarrow{}$ 0$^{+}$ becomes constant as x $\xrightarrow{}$ $\infty$ ) to the last three points of the cumulative j profile. We then, take the ratio $r_{k}$ (k being baryons/gas) of $j_{k}$ at last measured point to the maximum value that $\mathcal{P}$ would have if extrapolated to infinity. This ratio, (i.e. the convergence factor), is calculated for all the j$_{\rm bar}$ and j$_{\rm gas}$ profiles and listed for all the galaxies in Tables \ref{Table4}, \ref{Table5} and \ref{Table6}. Only 2 of the early-type galaxies (NGC 3626 and NGC6798) and 1 of the ultra-diffuse galaxies have a convergence factor $<$ 0.7.

  Several angular momentum studies, e.g.  \citet{murugeshan20}, \citet{elson17} do not check for convergence. Disk galaxies show a range of rotation curve shapes with low mass galaxies typically having rising rotation curves. However, in general, dwarf galaxies show a large diversity of rotation curve shapes, even at fixed maximum rotation speed \citep[e.g.][]{oman15}. Restricting the analysis to a set of galaxies that meet a user defined convergence criterion may lead to a non representative sample, particularly for the galaxies with lower baryonic masses.  For studies where one is considering the specific angular momentum of the galaxy disk, (as opposed to making inferences about the halo), it is not clear that a restriction of the sample to those whose specific angular momentum curves  meet a user defined convergence is essential.

\begin{figure}
  \centering
  \includegraphics[width=1.05\linewidth]{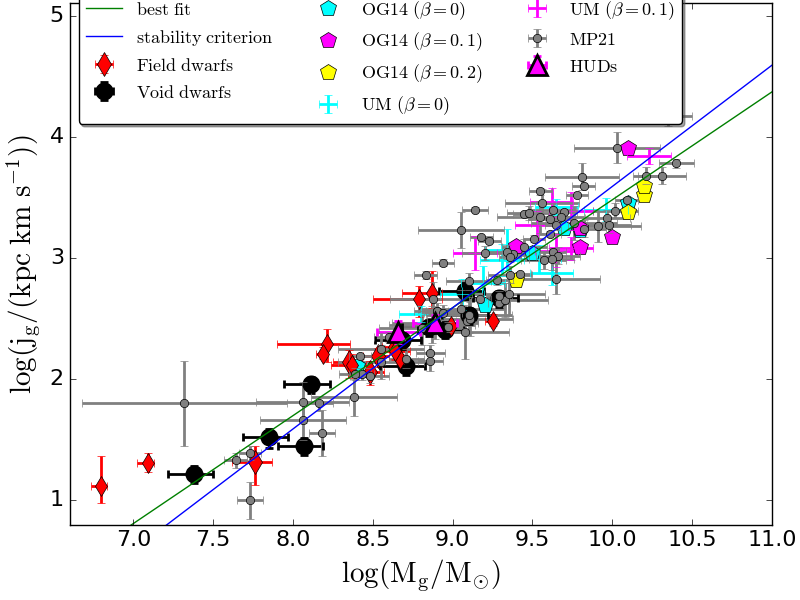}
  \caption{Log \jg -- log \mg  of 2 HUDs (magenta triangles) from this work, 11 dwarfs residing in Lynx-Cancer void (black circles) from \citet{kurapati18} , 17 field dwarfs (red diamonds from \citet{butler17} and \citet{chowdhury17}), 16 spirals residing in the UMa cluster (plus symbols) from this work, 15 spirals from \citet{obreschkow14} (pentagons), and 84 galaxies from \citet{mancera21} (grey circles). The cyan, magenta, and yellow symbols (plus symbols and pentagons) indicate spiral galaxies with bulge fractions, $\beta$ = 0, $\beta$ = 0.1, and $\beta$ = 0.2 respectively.
  The green line indicates the best fit  for the \jg -- \mg relation using the linear regression. The blue line represents the  stability criterion (j$_{\rm HI}$ x 11 km s$^{-1}$/(G M$_{\rm HI}$) $\sim$ 1) from \citet{romeo20}}
  \label{fig:jgvsmg}
\end{figure}

\section{Results}
\label{results}
\subsection{Specific angular momentum of the \HI disk}
\label{ssec:hi_sam}

As discussed in the introduction, recent simulations indicate that the angular momentum of galaxies is governed more by evolutionary processes and  the physics of the baryonic material than by the the initial spin of the `parent' dark matter halo. Numerical simulations also support the picture that star formation occurs in marginally stable gas disks, where inflows provide both the gas required for the star formation, as well as energy to keep the disks marginally stable against gravitational instability (e.g. \cite{dekel13}). Models of disk galaxies where the disks are critically stable also appear to provide a reasonable description of the specific angular momentum-mass relation  of the gas disks of local galaxies \citep[e.g.][]{zasov74,zasov17,kurapati18,romeo20}. Observational constraints on models where the gas mass and angular momentum of \HI disks are set by the requirement that the  disk be critically stable have been presented by e.g. \citet{obreschkow16, zasov17}. It would hence be  interesting to look at how well such models fit  the \HI specific angular momentum  (\jg) and the \HI mass (\mg) for our heterogeneous sample of galaxies.

We show in Fig.~\ref{fig:jgvsmg} the \HI specfic angular momentum and \HI gas mass relation for our sample galaxies along with other samples taken from the literature (details are given in the figure caption). Interestingly,  the galaxies show a smooth \jg -- \mg\ relation across the entire mass range. We estimate the best fitting linear regression between \jg\ and \mg\ by using the BCES algorithm \citep{akritas96} that accounts for errors on both \jg\ and \mg\ as well as for the fact that these errors are correlated. The best-fit \jg\ -\mg\ relation is given by, \jg\ = q\mg $^{\alpha}$ with $\alpha$ = 0.89 $\pm$ 0.05 and log$_{\rm 10}$ q = -5.42 $\pm$ 0.41.  We do not include early--type  galaxies while fitting for the \jg\ -- \mg\ relation since these galaxies are peculiar in that the gas is counter-rotating or misaligned with the stellar kinematics. We discuss these galaxies further below. The  existence of such a relation between the \HI mass and the \HI specific angular momentum (suggested earlier also by e.g. \cite{zasov74}) is interesting in itself, and we return to this in Sec.~\ref{sec:discussion}. We also note that the slope that we find for the \jg\ -- \mg\ relation is significantly different from that seen for the \js\ -- \ms\ relation (i.e. $0.55 \pm 0.02$) by \cite{posti19}. The slope we get is slightly smaller than (but statistically consistent with)  the value of $ 1.02 \pm 0.04$ recently reported by \citet{mancera21}.  We return to a discussion of this in Sec.~\ref{sec:discussion}. Here we focus on how the relation that we find compares with expectations from the critically stable disk model.

The commonly used criterion for stability is the Toomre Q parameter. Strictly speaking, the Toomre criterion is for a thin single component disk, while at least some of the galaxies that we are analysing have significant stellar masses as well as a non negligible disk thickness. Issues associated with applying the Toomre-Q criterion to such galaxies have been discussed in detail in  \citet{zasov17} and \citet{romeo20}. \citet{romeo20} in particular takes a phenomenological approach, and suggests that the different components (molecular, atomic, stellar) of galaxy disks follow scaling laws derived from the relation

$$ {j_i \hat{\sigma_i} \over G M_i} \approx 1 $$

where $\hat{\sigma_i} = \bar{\sigma_i}/A_i$, with $\bar{\sigma_i}$ being the disk averaged velocity dispersion of component $i$ (\HI, H$_2$, or stars), and $A_i$ is the mass averaged Toomre-Q parameter. The value of $\hat{\sigma_i}$ is not derived from fundamental principals, but is instead determined by a fit to observational data. \citet{romeo20} uses data from \citet{leroy08} (which overlaps with the \citet{obreschkow14} sample) to determine the value for $\hat{\sigma_i}$. For the \HI component the derived value is 11~km/s. We show this empirical relation, log(j) = log(M)-6.41 (with $\hat{\sigma_i} = 11$~km/s) in Fig.~\ref{fig:jgvsmg}. As can be seen, it provides a good fit to the data,  the majority of  which is independent from the sample used by \citet{romeo20}.  This further supports the idea that the structure of the \HI disks in galaxies settles to a critical density, as would happen in models where the star formation rate (and hence mass outflow rate) is driven by the mass inflow rate. 

As discussed in the introduction, simulations also show that spheroidal stellar distributions with extended counter-rotating gas disks could arise in systems where there has been a recent accretion of gas whose spin is not aligned with that of the parent galaxy \citep[e.g.][]{kretschmer20}. In the context of such models it would be interesting to examine the position of extremely gas rich early--type  galaxies in the $\jg\ - \mg$ plane.  As can be seen from Table~1, for many  of the galaxies in our sample the gas is counter-rotating or misaligned  with the stellar kinematics, consistent with results from simulations. Figure \ref{fig:S0} shows the location of early--type  galaxies, as well as all our other sample galaxies in the \jg\ -- \mg\ plane. The green solid line indicates the best fit line using the linear regression for all the galaxies except the early--type  galaxies and green dotted lines indicate 3$\sigma$ vertical scatter. As can be seen, most of the early--type  galaxies lie above the 3$\sigma$ scatter, i.e. their \HI disks have significantly higher specific angular momentum as compared to other galaxies with the same \HI mass. A Kolmogorov--Smirnov (KS) test\citep[][\url{https://github.com/syrte/ndtest/blob/master/ndtest.py}]{fasano87}  gives a probability  of 7 $\times$ 10$^{-4}$ for the early--type  galaxies being  drawn from the same distribution as the other galaxies.

   Two of these early-type galaxies (NGC3626 and NGC6798, marked with a circle around the blue open triangles in Fig \ref{fig:S0}) do not satisfy the convergence criterion. We note that they  do not appear to have discrepant \jg\ as that of the early-type galaxies with converged specific angular momentum curves. If the lack of convergence means that these two galaxies have a higher specific angular momentum than measured, it does not affect the result that the  early–type galaxies in our sample have significantly higher \jg\ than expected from the \jg - \mg\ power law. Our observations hence  support the picture in which  these gas rich early-type galaxies have acquired the gas via a recent inflow of high, but misaligned sAM gas. The elevated specific angular momentum is consistent with the  decreased star formation efficiency of early--type  galaxies since higher angular momentum  stabilizes the gas against Jeans instabilities and subsequent star formation \citep{toomre64, obreschkow16}.

 We also show the scaling relation between the atomic gas fraction  (f$_{\rm atm}$) and the stability parameter (q) for all the galaxies in our sample in Figure \ref{fig:gasfrac_q} for which these parameters are available.  The dotted line shows the relation expected from \citet{obreschkow16}. 
We find that all the galaxies in our sample largely follow the relation consistently. We do not show early--type galaxies in this figure since we can not measure their baryonic specific angular momentum as the gas does not in general co-rotate with the stars for these galaxies.  The dwarf galaxies lie in the saturated region (f$_{\rm atm} \sim $ 1).  We also show the H{\sc i} excess (higher atomic gas fraction) galaxies from \citet{lutz18} for comparison; they follow the relation between the atomic to baryonic mass fraction and the global stability parameter given by \citet{obreschkow16}. 

\begin{figure}
\centering
\includegraphics[width=1.05\linewidth]{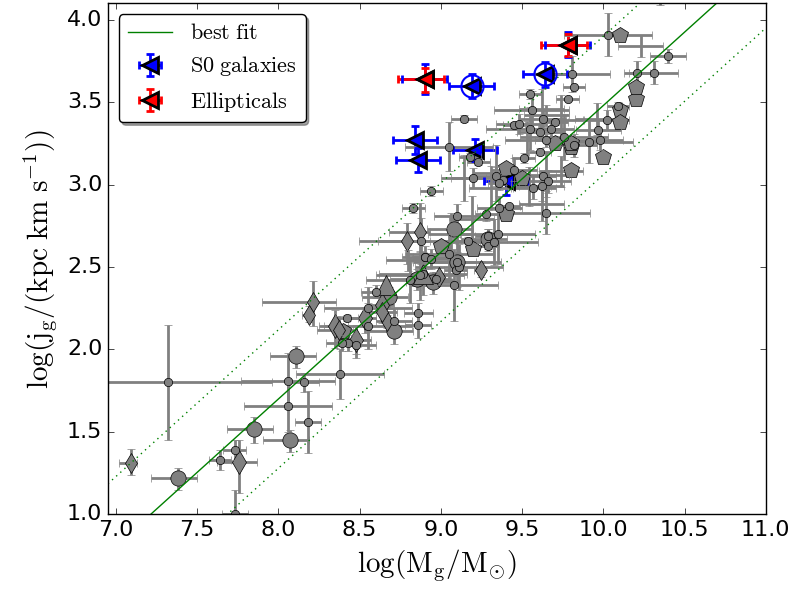}
\caption{ The location of early--type  galaxies with large regular \HI disks (blue open triangles) and the other galaxies (symbols are the same as in Figure~\ref{fig:jgvsmg}) in the  \jg-\mg plane. The green solid line indicates the best fit line using the linear regression for all the galaxies except early--type  galaxies. The green dotted lines indicates 3$\sigma$ scatter around the relation. The early-type galaxies that do not follow the convergence criterion are marked with a circle around the blue open triangles. The \HI disks in gas rich early--type  galaxies have high specific angular momentum compared to other galaxies with similar \HI mass.}
\label{fig:S0}
\end{figure}
\begin{figure}
  \centering
  \includegraphics[width=1.05\linewidth]{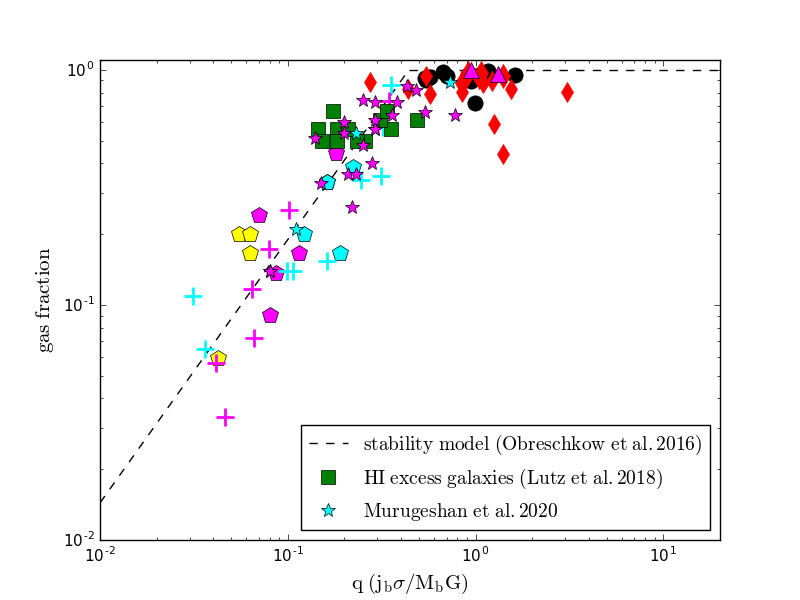}
  \caption{The atomic gas mass to baryonic mass fraction versus the global stability parameter q for all the galaxies in our sample (symbols are the same as in Figure~\ref{fig:jgvsmg}). We also show galaxies from \citet{murugeshan20} (cyan stars ($\beta$ = 0) and magenta stars ($\beta$ = 0.1)) and the H{\sc i} excess galaxies from \citet{lutz18} (green squares). These galaxies were not included in the earlier plots for the \jg-\mg relation because \jg is not available for these galaxies. 
  We do not show the galaxies from \citet{mancera21} as q values were not presented. The dotted line is the relation predicted by \citet{obreschkow16}. }
  \label{fig:gasfrac_q}
\end{figure}

\subsection{Baryonic Specific Angular Momentum}

\begin{figure}
  \centering
  \includegraphics[width=1.05\linewidth]{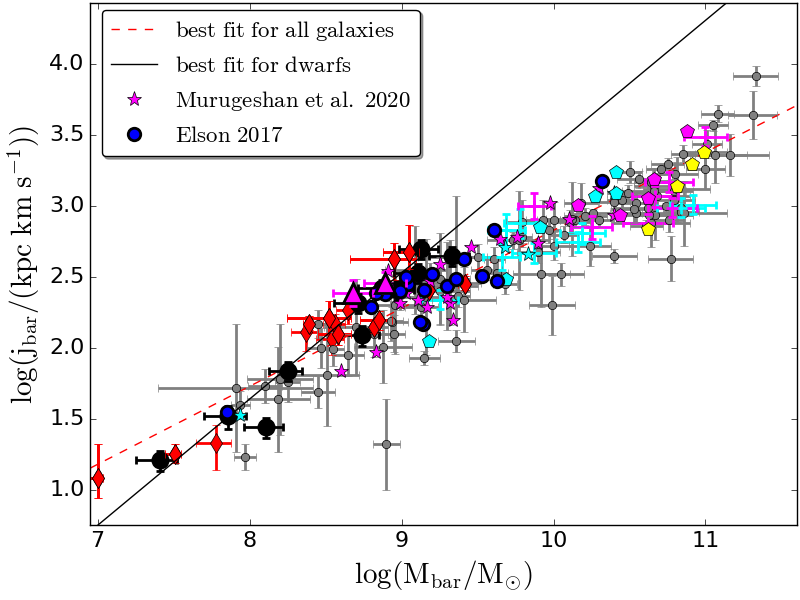}
  \caption{The log \jb -log \mb plane. The black solid line is the best fit line for the dwarf galaxies from \citet{kurapati18} and the red dashed line is the best fit line for all the galaxies. The symbols are the same as in Figure \ref{fig:jgvsmg}. The blue circles represent galaxies (only the galaxies with reliable rotation curves) from \citet{elson17}.}
  
  \label{fig:all_bar}
\end{figure}

Several earlier studies \citep[e.g.][]{obreschkow14, kurapati18,murugeshan20,mancera21}  have studied  the correlation between the specific angular  momentum and mass of the baryonic (i.e. gas + stars) component of galaxies. In particular \citet{kurapati18} investigated the relation considering the baryonic mass for a sample of gas rich dwarf galaxies in a range of environments, using the bulgeless spiral relation found  by \citet{obreschkow14} for comparison. They concluded that the baryonic specific angular momentum of dwarfs was higher as compared to the extrapolation of the trend seen for higher mass bulge-less spirals. The increase in specific angular momentum was found to set in at a mass threshold of $ \sim$ 10$^{9}$ M$_{\odot}$. However, more recent studies (e.g. \cite{mancera21}) find that the \jb-\mb relation is well fit by a single power law.

 Fig.~\ref{fig:all_bar} shows the relation between the baryonic specific angular momentum and the baryonic mass for the UMa sample of spiral galaxies from this work, along with the galaxies in \citet{obreschkow14, butler17, chowdhury17, elson17, kurapati18, murugeshan20, mancera21} We do not use the early--type  galaxies from the current work this analysis since, as discussed above for these galaxies the gas does not in general co-rotate with the stars. If fit a single power law to our sample (red dashed line in Fig. \ref{fig:all_bar}), we obtain a slope 0.55 $\pm$ 0.02 in good agreement with the slope measured by \citet{mancera21}. However, all the galaxies having \mb $\lesssim$ 10$^{8}$ M$_{\odot}$ lie significantly below the relation while the galaxies having   10$^{8}$ M$_{\odot} \lesssim $ \mb $\lesssim$ 10$^{9}$ M$_{\odot}$ are consistently above the relation. In this context, we note that the \citet{mancera21} sample does not contain galaxies with baryonic mass lower than 10$^{8}$M$_{\odot}$. This suggests that there may be some curvature in the the \jb --\mb\ relation, and it may not be best described by a single unbroken power law.

In order to test how well the \jb -- \mb relation is  described a single unbroken power law, we estimate the best fit linear regression between \jb and \mb independently for two separate mass ranges, viz. galaxies with \mb $>$ 10$^{9}$ M$_{\odot}$ and for the galaxies with \mb $<$ 10$^{9}$ M$_{\odot}$.  We find significantly different slopes for the \jb -- \mb relation, with a slope of 0.80 $\pm$ 0.05 for low mass galaxies and a slope of 0.52 $\pm$ 0.02 for high mass galaxies. This suggests that a single power law is not the best description of the data. On the other hand, the slopes of \jg - \mg relation match within the error bars for low mass galaxies ($\sim$ 0.85 $\pm$ 0.05) and high mass galaxies ($\sim$ 0.93 $\pm$ 0.07), suggesting that a single power law is an adequate description.

We also use the ``Chow test" \citep{chow60} to test whether the \jb -- \mb relation is well fit by a single unbroken power law. The test  checks for the statistical significance of postulated change in slope in a regression. We run the Chow test on both the \jb -- \mb  and \jg - \mg relations (\url{https://github.com/jtloong/chow_test}) to test the significance of a difference in slope between the high and low mass ends, with the nominal dividing point taken as  10$^{9}$ M$_{\odot}$. As per the Chow test, the null hypothesis that the baryonic j-M relation follows a unbroken power law is rejected at a high significance, with a p value of 5.5 $\times$ 10$^{-5}$. This indicates that there is a significant change in slope, with a broken power law providing a better fit. We note that the test does not independently identify a sharp break-point; a curved relation  with a gradual change in slope as one moves from lower to higher baryonic masses would also be better fit by a broken power law rather than a single one. In contrast to this, we find a fairly large p value (0.2) for the gas j-M relation which indicates that it is adequately fit by a single power law. We note that the number of galaxies with low baryonic masses in the sample is still quite modest, and it is important to increase the sample size at the low mass end. We return to a discussion of the possible reasons for the change in slope seen in the $\jb\ - \mb$ relation in the next section. }

Finally, we  draw attention to one more feature in Fig.~\ref{fig:all_bar}, the location of the ultra-diffuse dwarfs. Ultra-diffuse dwarf galaxies (UDGs) are a recently discovered population of very low surface brightness galaxies, which have stellar masses of dwarfs, but have sizes as large as L$_{\ast}$ galaxies \citep{dokkum15}. The majority of the UDGs are red and appear more common in clusters. However, galaxies with extreme ratios of stellar mass to stellar scale length were also discovered in H{\sc i} surveys of isolated environments (e.g. Leisman et al. 2017). These galaxies are called H{\sc i}--bearing ultra diffuse galaxies (HUDs), which are bluer and have significant reservoirs of H{\sc i} as compared to the UDGs in the clusters. Several formation scenarios have been proposed to explain the formation of the UDGs. Some authors propose a scenario in which UDGs are failed Milky Way like galaxies which lost their gas after forming the first stars \citep{dokkum15} and others argue that they are genuinely dwarf galaxies residing in high spin parameter halos \citep{amorisco16}. These formation scenarios can be tested by measuring the angular momentum of UDGs and comparing them to the normal dwarf galaxies. Figure \ref{fig:all_bar} shows the baryonic specific angular momentum versus the baryonic mass of H{\sc i}--bearing UDGs (magenta triangles) and normal dwarf galaxies of similar masses. We find that the specific angular momentum of H{\sc i}--bearing UDGs is similar to other normal dwarf galaxies at a given mass. This is in contrast with the formation scenarios that predict that UDGs are genuinely dwarf galaxies residing in high spin haloes. We note that our result is in contradiction with a recent analysis by \citet{mancera21}, who find that ultra diffuse galaxies have higher-than-average {\it stellar} specific angular momentum, but have circular velocities much lower than galaxies with similar baryonic mass. They assume flat rotation curves and exponential light curves and approximate the stellar specific angular momentum to be the product of circular velocity and the stellar disk scale length since they lack the resolution to integrate angular momentum over the entire disk. This, along with the very small size of our sample, make it important to confirm these results with high resolution observations of a larger sample.

\section{Discussion}
\label{sec:discussion}

As we saw above both the \jg\ -- \mg\ relation (this work and \citet{mancera21}) as well as  the \js\ -- \ms\ relation \citep{posti18} follow an unbroken power law  over the mass range of 10$^{7}$ -- 10$^{10.5}$ M$_{\odot}$. However their slopes are very different, viz.  with slopes of $0.89 \pm 0.05$  (this work) and $0.55 \pm 0.02$  \citep{posti18} respectively. In contrast to these relations, the \jb\ - \mb\ relation is better fit by a broken power law, with gas rich low mass galaxies having higher specific angular momentum compared to the extrapolated trend from high mass galaxies. We test the scenario that this increase in baryonic specific angular momentum essentially reflects the different relative contributions of the stellar and gas angular momenta to the total baryonic angular momentum following from the different slopes of stellar and gas j--M relations.

\begin{figure}
\centering
\includegraphics[width=1.05\linewidth]{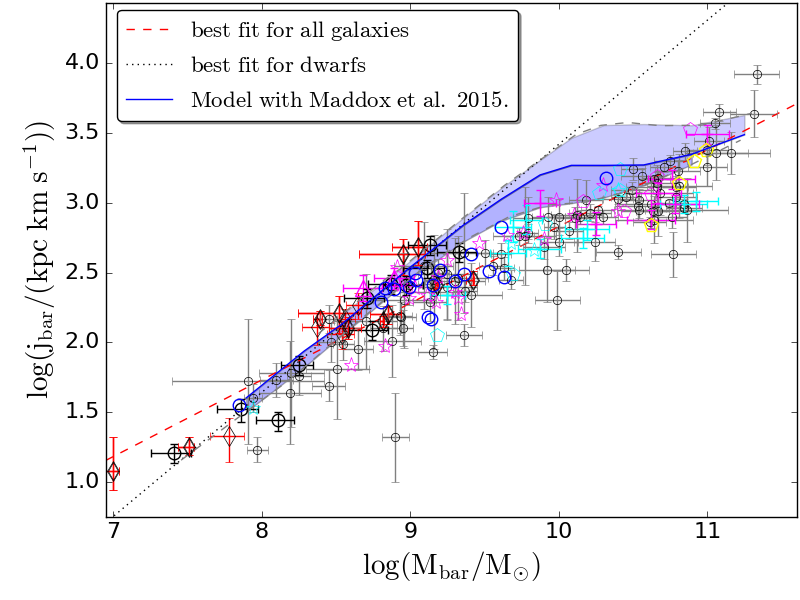}
\caption{The log \jb -log \mb plane. The black dotted line is the best fit line for the dwarf galaxies from \citet{kurapati18} and the red dashed line is the best fit line for all the galaxies. The symbols are the same as in previous figure. The model is shown as blue shaded region }
\label{fig:model}
\end{figure}

 We calculate the general expected trend  of \jb\ with \mb\ by using the observed  M$_{\rm g}$ - M$_{\rm s}$ and \jg - \mg\ and j$_{\rm s}$ -- M$_{\rm s}$ relations as well as the relation between M$_{\rm HI}$ and  M$_{\rm s}$ from \citet{maddox15}.  The predicted \jb\ - \mb\ curve from these relations is shown in Fig. \ref{fig:model}, where  we have included the contribution of helium to the gas mass by multiplying M$_{\rm HI}$ by a factor of 1.35. For the predicted relation we have used the median H{\sc i} mass values and their 1$\sigma$ uncertainties derived for galaxies in the \citet{maddox15} sample for which  SDSS spectra are available.  The blue solid curve indicates the expected \jb - \mb relation and the blue shaded region indicates the 1$\sigma$ scatter. The relation obtained if use the relation from \citet{maddox15} derived from the sample that includes galaxies for which no SDSS spectrum is available matches the blue curve within the 1$\sigma$ scatter. As can be seen, the model curve is in broad qualitative agreement with the data showing a change in slope around $10^{9}\msun$, although the $\jb$ values in this mass range are somewhat larger than the observed points. We note that the actual gas fractions of the sample being used would shift the location of the points in the \jb --\mb\ plane. However, the broad qualitative agreement between this simple model and the data supports the idea that the different slopes obtained at the low and high mass ends of the $\jb\ - \mb$ relation  is a consequence of the changing contribution of the stars and gas to the total angular momentum at the high and low $\mb$ ends. 

   It is also interesting to ask how the power law relations between the specific angular momentum and the mass fit in  with the other known scaling relations? To address this question we consider a simple model galaxy where the stars and gas have exponential surface brightness profiles  with scale lengths r$_{\rm s}$ and r$_{\rm g}$ as well as a flat rotation curve with V(r) = V$_{\rm c}$. This gives  j $ \propto$ V$_{\rm c}$r.  From this expression we calculate the expected slope ($\alpha_{\rm s}$) for the \js\--\ms\ relation by using the stellar mass size relation, r$_{\rm s}$ $\propto$ M$_{\rm s}^{0.28}$ \citep{bell01} and the stellar TF relation, V$_{\rm c}$ $\propto$ M$_{\rm s}^{0.31}$ \citep{bloom17}, which gives $\alpha_{\rm s}$ $\sim$ 0.28+0.31 = 0.59 which compares well  with the observed slope of $\sim$ 0.55 $\pm$ 0.02.  For the gas rich galaxies we use the baryonic TF relation, V$_{\rm c} \propto $ M$_{\rm bar}^{0.33}$ \citep{ponomareva18} and assume that the baryons are dominated by gas for the dwarf galaxies, which gives V$_{\rm c} \propto $ M$_{\rm g}^{0.33}$.  Observations suggest that the H{\sc i} mass and radius are related as r$_{\rm g} \propto$ M$_{\rm g}^{0.51}$ \citep{wang16}. This also arises from  simulations if we assume a exponential H{\sc i} disk, with the inner radius and outer radius being set by molecular transition and observational threshold for detecting H{\sc i} \citep{obreschkow09}.  
The combination of baryonic TF relation and the gas mass-size relation gives the slope ($\alpha_{\rm g}$) for the \jg\ -- \mg\ relation, $\alpha_{\rm g}$ $\sim$ 0.51+0.33 = 0.84, in excellent agreement with what we observe. For less gas rich galaxies, if we  combine the stellar TF relation V$_{\rm c}$ $\propto$ M$_{\rm s}^{0.31}$ from \citet{bloom17} and the gas fraction relation \mg / \ms = \ms $^{-0.53}$ from \citet{romeo20}, we find V$_{\rm c} \propto $ M$_{\rm g}^{0.66}$.  and $\alpha_{\rm g}$ $\sim$ 0.66+0.51 = 1.17 in reasonable agreement with what we observe.  We cannot determine the predicted values of the intercepts, since they depend on the ratio of the size of the HI disk  to its scale length, which is not well measured.


\section{Summary and Conclusions}
\label{summary}

We have examined the relation between the \HI\ mass and the \HI\ specific angular momentum for a sample of galaxies with well measured rotation curves that spans a range of morphology. Combining this with data from the literature, we find that the relation is well fit by an unbroken power law over the mass range $10^{7} - 10^{10.5}$~\msun. The scatter in this relation is small, and this supports earlier suggestions that the relationship between the angular momentum and mass is a fundamental scaling relation for galaxies. The slope that we measure for the power law $0.89 \pm 0.05$ is also significantly different from that expected for dark matter haloes from tidal torquing models which predict $j \propto M^{2/3}$, as well as the observed slope of $0.55 \pm 0.02$ for the stellar component of disk galaxies.  The slope we get for the $\jg-\mg$\ relation is slightly smaller than (but statistically consistent with)  the value of $ 1.02 \pm 0.04$ recently reported by \citet{mancera21}. Our sample includes two \HI\ bearing ultra-diffuse galaxies, and we find that they lie along the relationship defined by the other galaxies in our sample. The only discrepant galaxies are early--type  galaxies with large well rotating disks. These are found to have excess angular momentum compared to that predicted by the relation. In all of these galaxies, the \HI\ disk also appears to be kinematically distinct from the stellar disk, indicating that it represents recently acquired material.  Overall, our observations leads us to suggest that the kinematics of the baryonic component of gas rich galaxies is more affected by cold flows and late stage mergers than by the spin parameter of the `parent' dark matter halo as suggested by recent numerical simulations. We also find that the $\jb-\mb$ relation is better fit by a broken power law than a single power law, indicating that there is some curvature in this relation. We suggest that the change in slope of the $\jb-\mb$ relation arises from the changing gas to stellar mass ratio with mass as well as the different slopes of the $\js-\ms$ and $\jg-\mg$ relations. 

\section*{Acknowledgements}

 This paper is based in part on observations taken with the GMRT. We thank the staff of the GMRT who made these observations possible. The GMRT is run by the National Centre for Radio Astrophysics of the Tata Institute of Fundamental Research. This work was supported by the DAE grant 12-R\&D-TFR-5.02-0700. MV acknowledges support by the Netherlands Foundation for Scientific Research (NWO) through VICI grant 016.130.338 and thanks the NCRA staff for their hospitality. We are grateful for insightful comments from one of the anonymous reviewers that helped improve this paper.




\bibliographystyle{mnras}
\bibliography{S0_angmom} 





\appendix

\section{Data}

\begin{figure}
\centering
\includegraphics[width=1.0\linewidth]{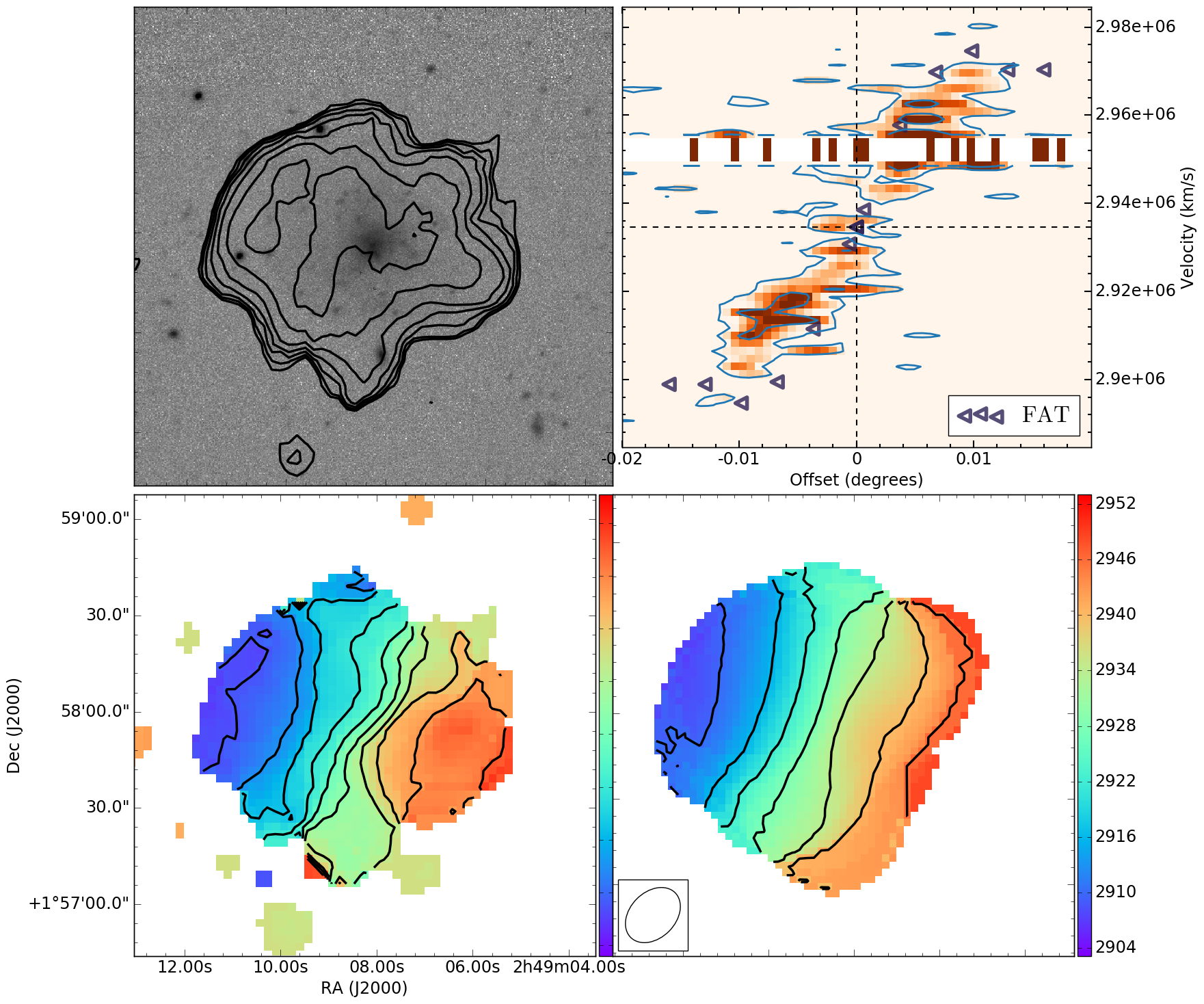}
\caption{H{\sc i} data and kinematics for the galaxy AGC121790. Top left: H{\sc i} distribution of the galaxy overlaid on the SDSS g–band data,Top right: Position velocity diagram (outer contour corresponds to 2$\sigma$) with the rotation curves overlaid on them. Bottom left: velocity field of data, and Bottom right:velocity field of the best fitting \fat model. velocity contours run from 2910 to 2946 km s$^{-1}$ with a spacing of 6 km s$^{-1}$.}
\label{fig:mom1}
\end{figure}

\begin{figure}
\centering
\includegraphics[width=1.0\linewidth]{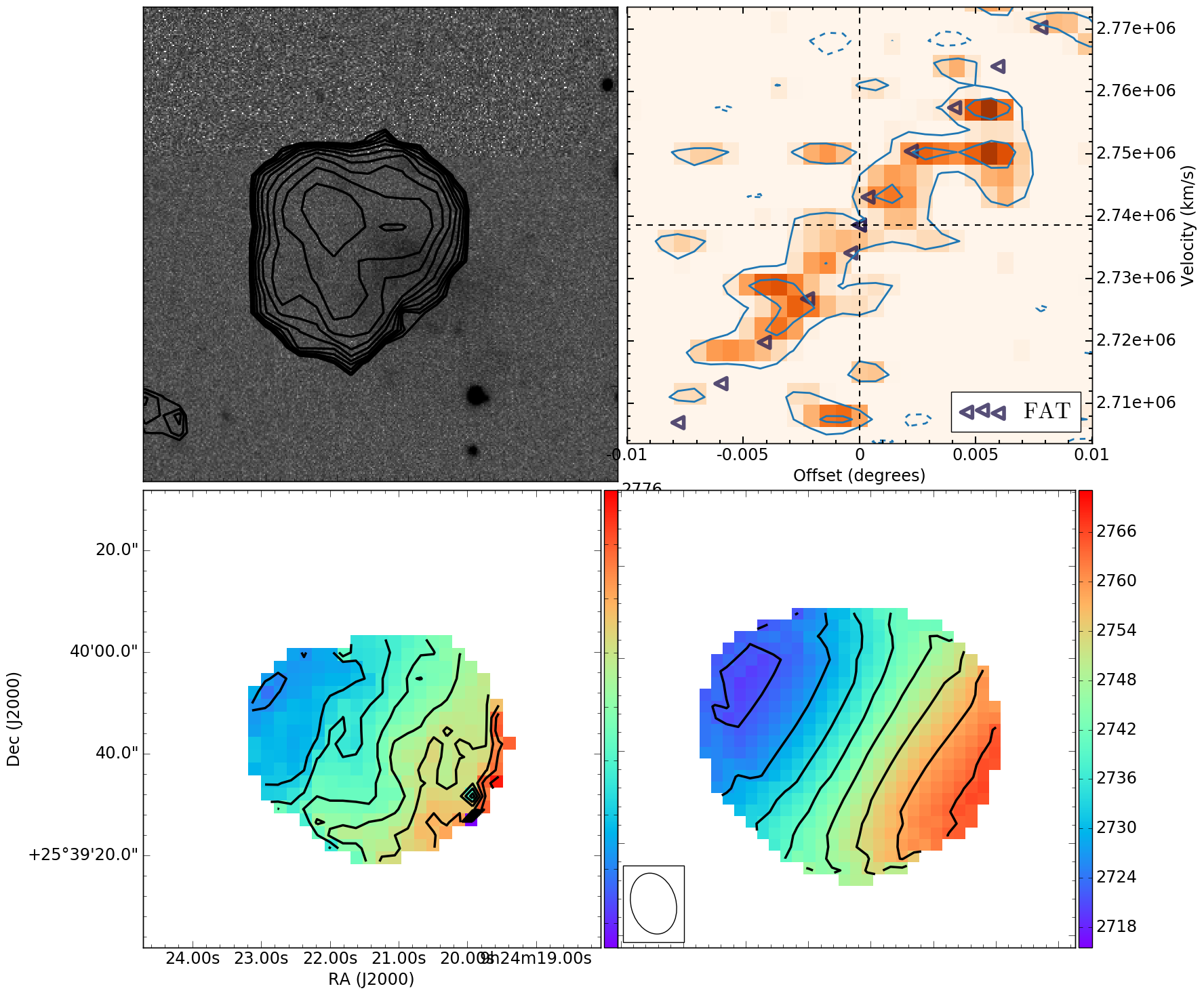}
\caption{H{\sc i} data and kinematics for the galaxy AGC749401. Top left: H{\sc i} distribution of the galaxy overlaid on the SDSS g–band data,Top right: Position velocity diagram  (outer contour corresponds to 2$\sigma$)with the rotation curves overlaid on them. Bottom left: velocity field of data, and Bottom right:velocity field of the best fitting \fat model. velocity contours run from 2712 to 2754 km s$^{-1}$ with a spacing of 6 km s$^{-1}$. }
\label{fig:mom2}
\end{figure}

\begin{figure}
\centering
\includegraphics[width=1.0\linewidth]{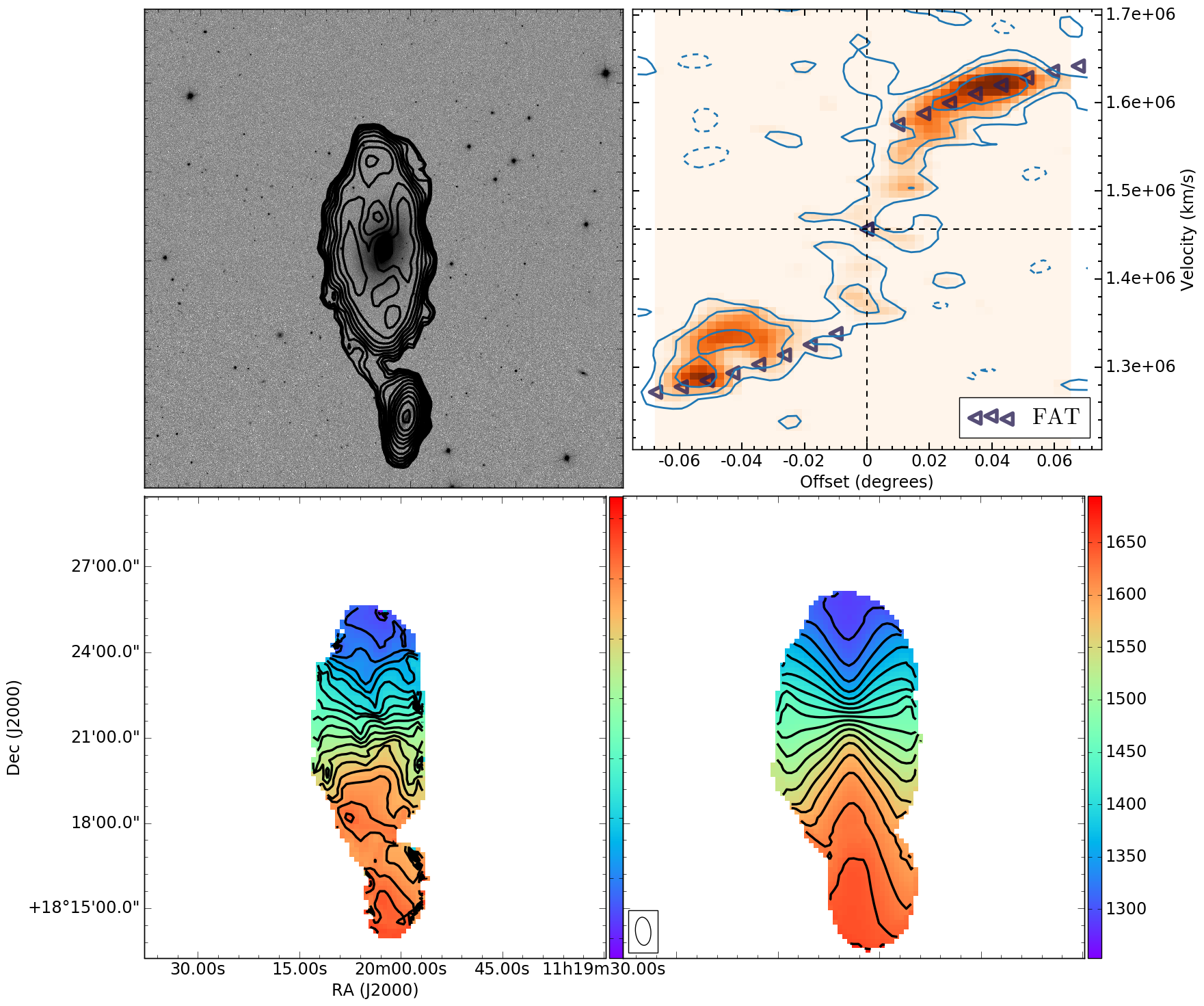}
\caption{H{\sc i} data and kinematics for the galaxy NGC3626. Top left: H{\sc i} distribution of the galaxy overlaid on the SDSS g–band data,Top right: Position velocity diagram (outer contour corresponds to 2$\sigma$) with the rotation curves overlaid on them. Bottom left: velocity field of data, and Bottom right:velocity field of the best fitting \fat model. Velocity contours run from 1300 to 1660 km s$^{-1}$ with a spacing of 20 km s$^{-1}$.}  
\label{fig:mom3}
\end{figure}

\begin{figure}
\centering
\includegraphics[width=1.0\linewidth]{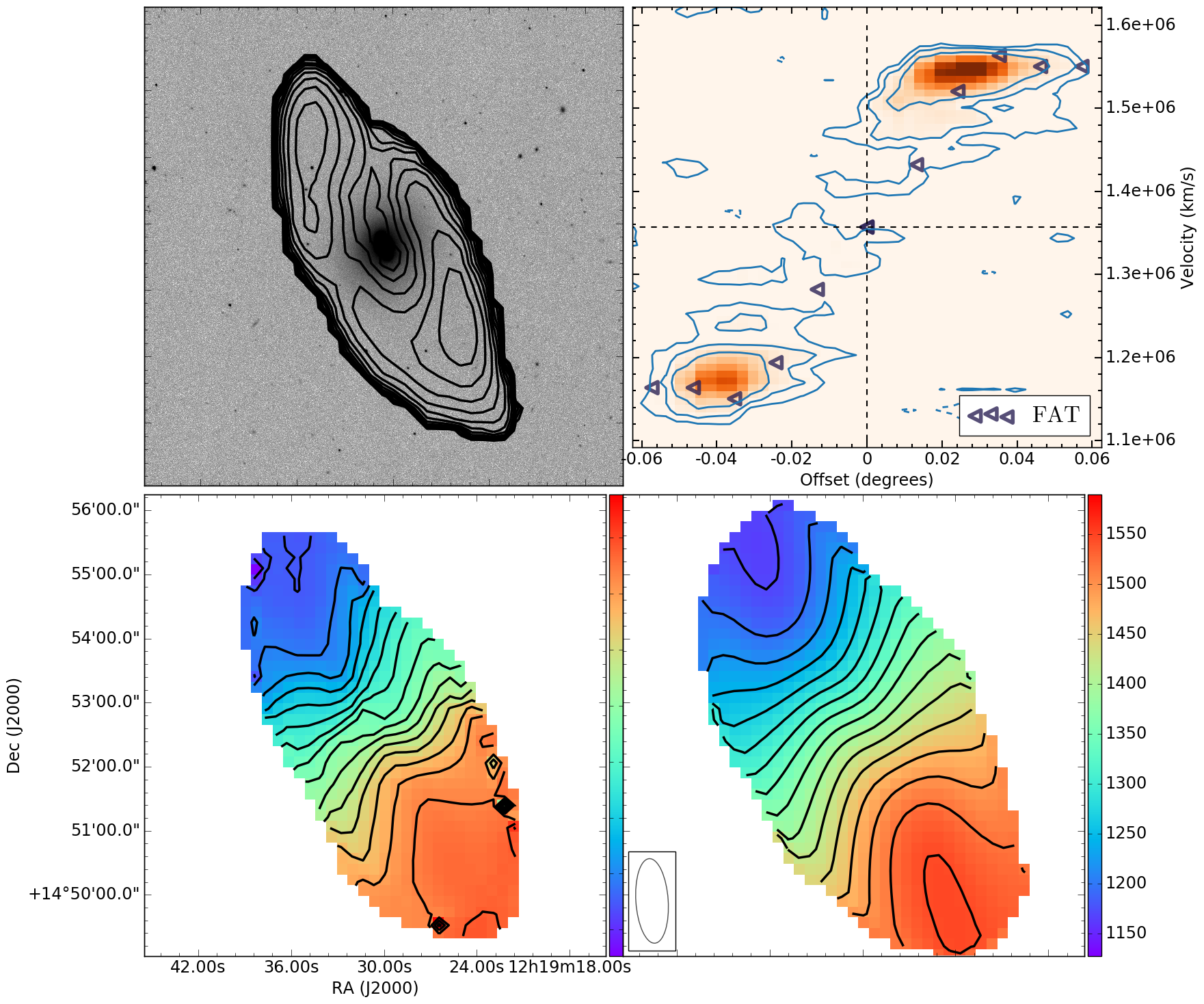}
\caption{H{\sc i} data and kinematics for the galaxy NGC4262. Top left: H{\sc i} distribution of the galaxy overlaid on the SDSS g–band data,Top right: Position velocity diagram (outer contour corresponds to 2$\sigma$) with the rotation curves overlaid on them. Bottom left: velocity field of data, and Bottom right:velocity field of the best fitting \fat model. Velocity contours run from 1145 to 1545 km s$^{-1}$ with a spacing of 25 km s$^{-1}$.}  
\label{fig:mom4}
\end{figure}

\begin{figure}
\centering
\includegraphics[width=1.0\linewidth]{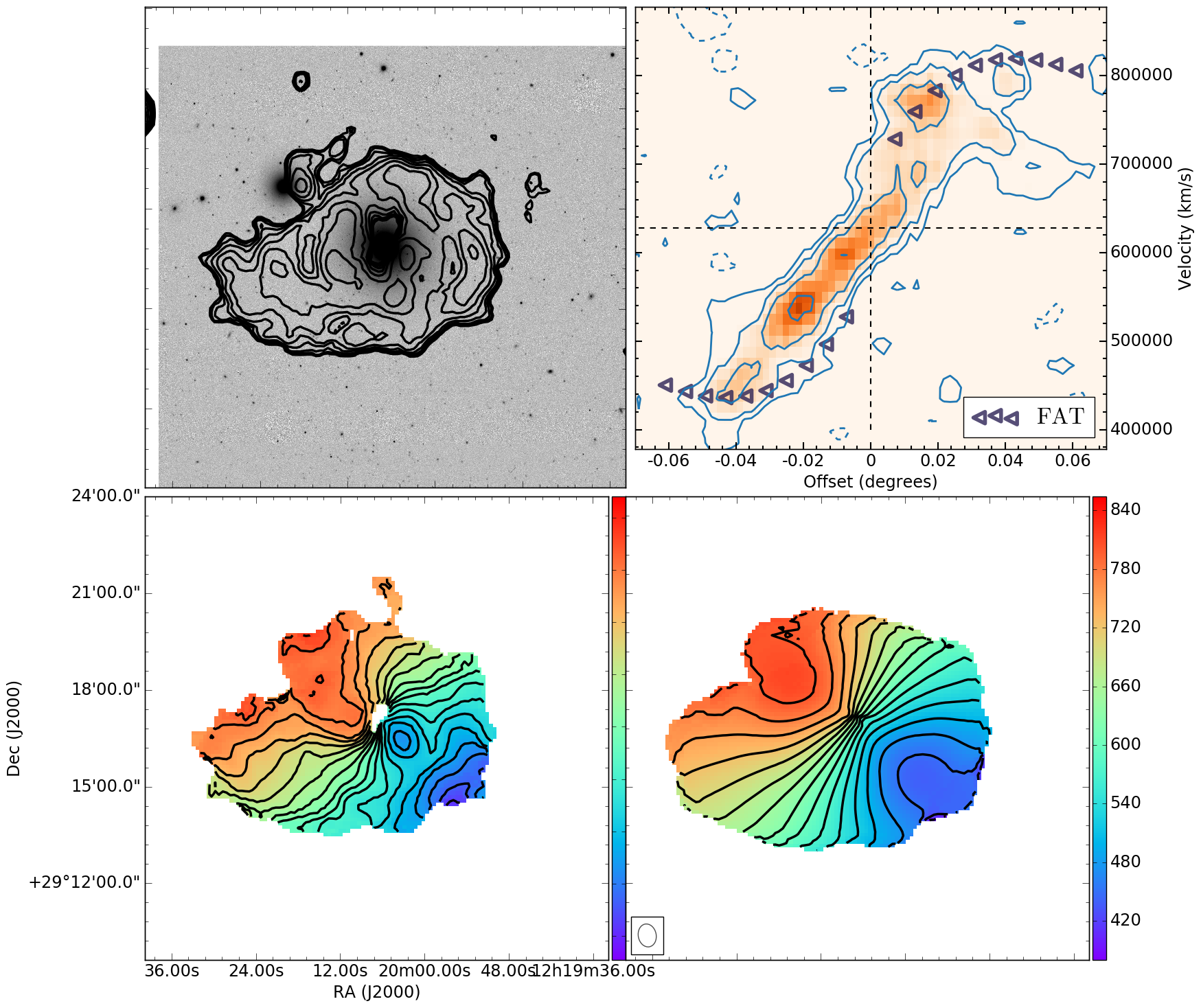}
\caption{H{\sc i} data and kinematics for the galaxy NGC4278. Top left: H{\sc i} distribution of the galaxy overlaid on the SDSS g–band data,Top right: Position velocity diagram (outer contour corresponds to 2$\sigma$) with the rotation curves overlaid on them. Bottom left: velocity field of data, and Bottom right:velocity field of the best fitting \fat model. Velocity contours run from 425 to 800 km s$^{-1}$ with a spacing of 25 km s$^{-1}$.}
\label{fig:mom5}
\end{figure}

\begin{figure}
\centering
\includegraphics[width=1.0\linewidth]{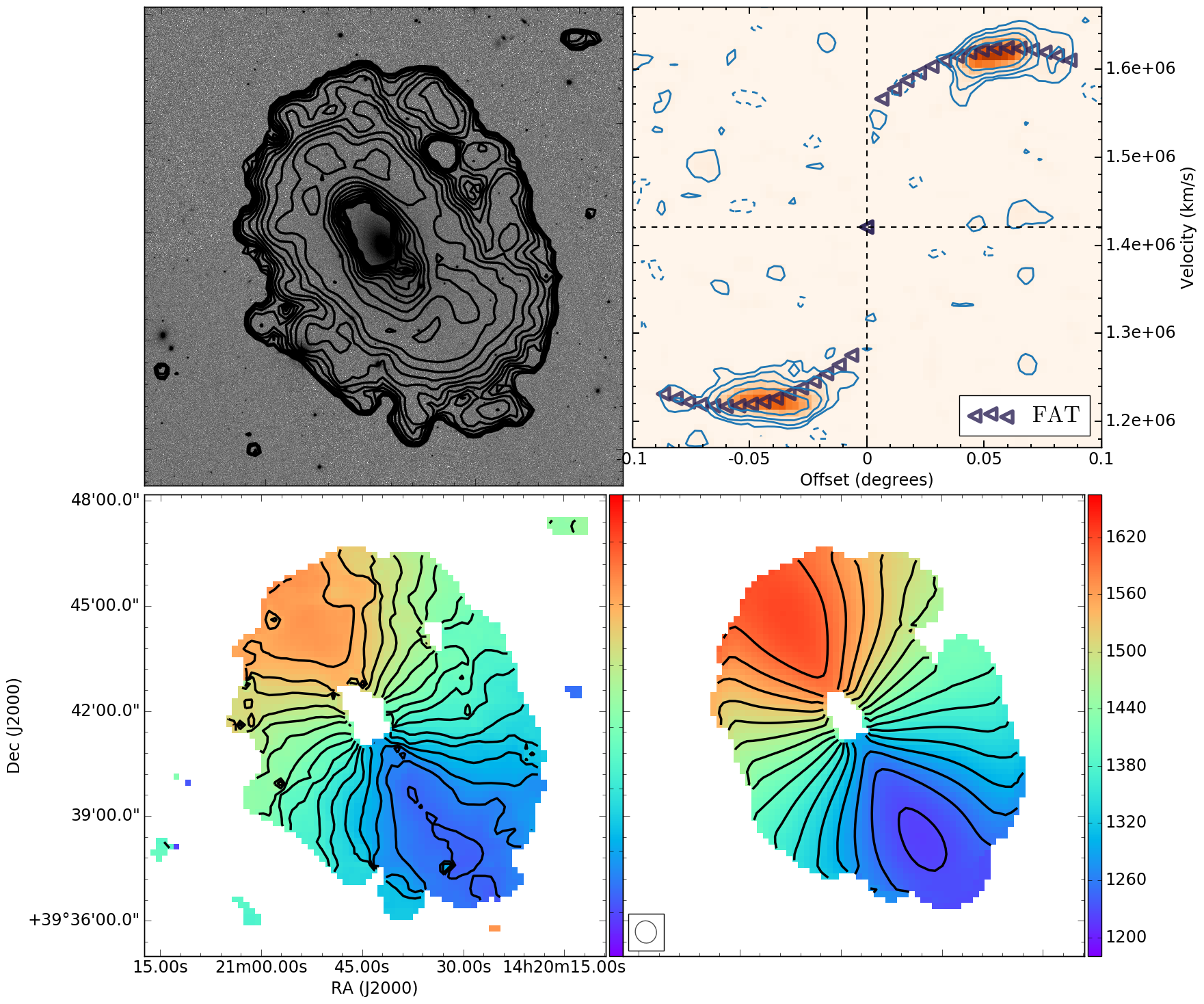}
\caption{H{\sc i} data and kinematics for the galaxy NGC5582. Top left: H{\sc i} distribution of the galaxy overlaid on the SDSS g–band data,Top right: Position velocity diagram (outer contour corresponds to 2$\sigma$) with the rotation curves overlaid on them. Bottom left: velocity field of data, and Bottom right:velocity field of the best fitting \fat model. Velocity contours run from 1225 to 1600 km s$^{-1}$ with a spacing of 25 km s$^{-1}$. }
\label{fig:mom6}
\end{figure}

\begin{figure}
\centering
\includegraphics[width=1.0\linewidth]{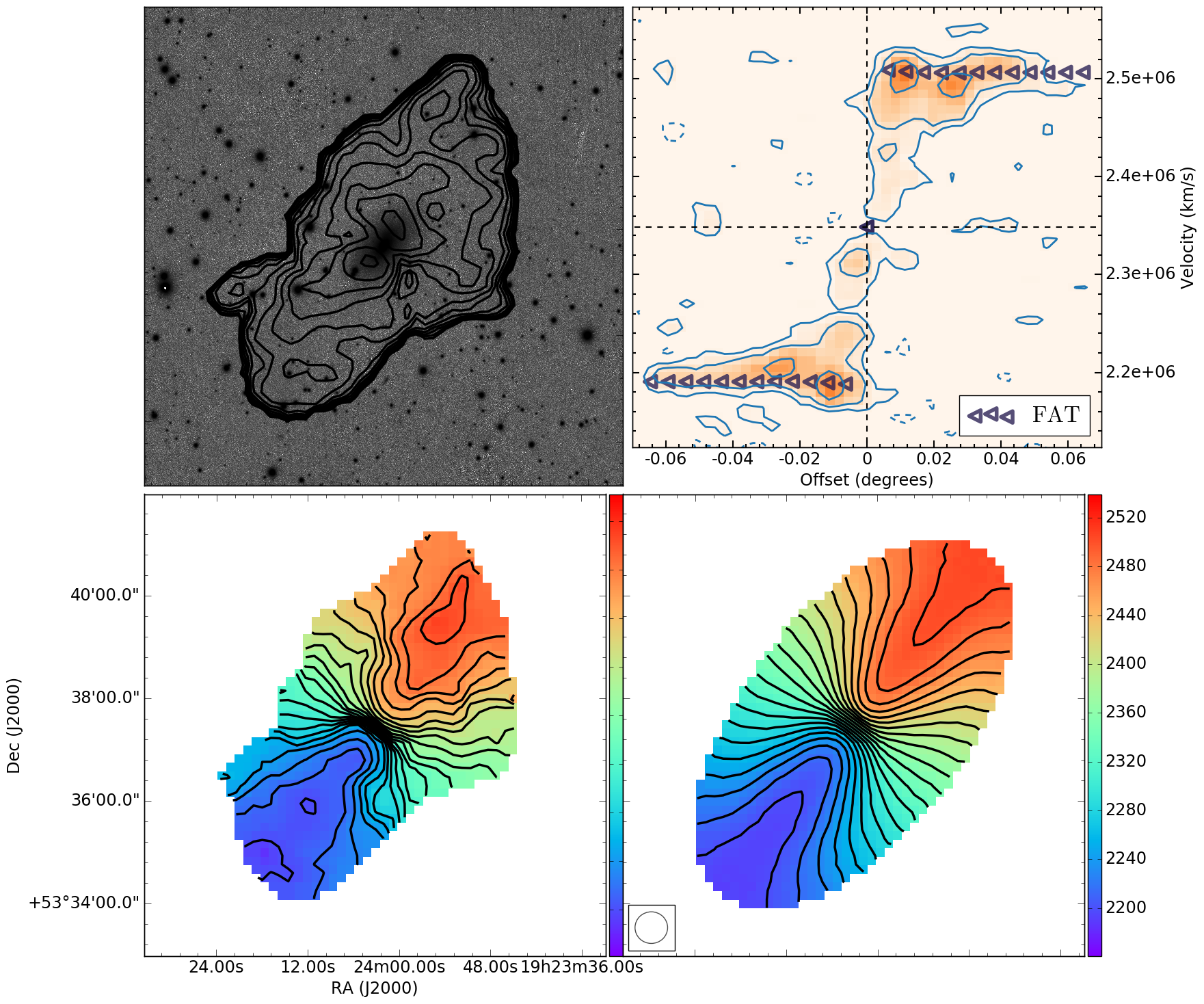}
\caption{H{\sc i} data and kinematics for the galaxy NGC6798. Top left: H{\sc i} distribution of the galaxy overlaid on the SDSS g–band data,Top right: Position velocity diagram (outer contour corresponds to 2$\sigma$) with the rotation curves overlaid on them. Bottom left: velocity field of data, and Bottom right:velocity field of the best fitting \fat model. Velocity contours run from 2200 to 2500 km s$^{-1}$ with a spacing of 15 km s$^{-1}$.}
\label{fig:mom7}
\end{figure}

\begin{figure}
\centering
\includegraphics[width=1.0\linewidth]{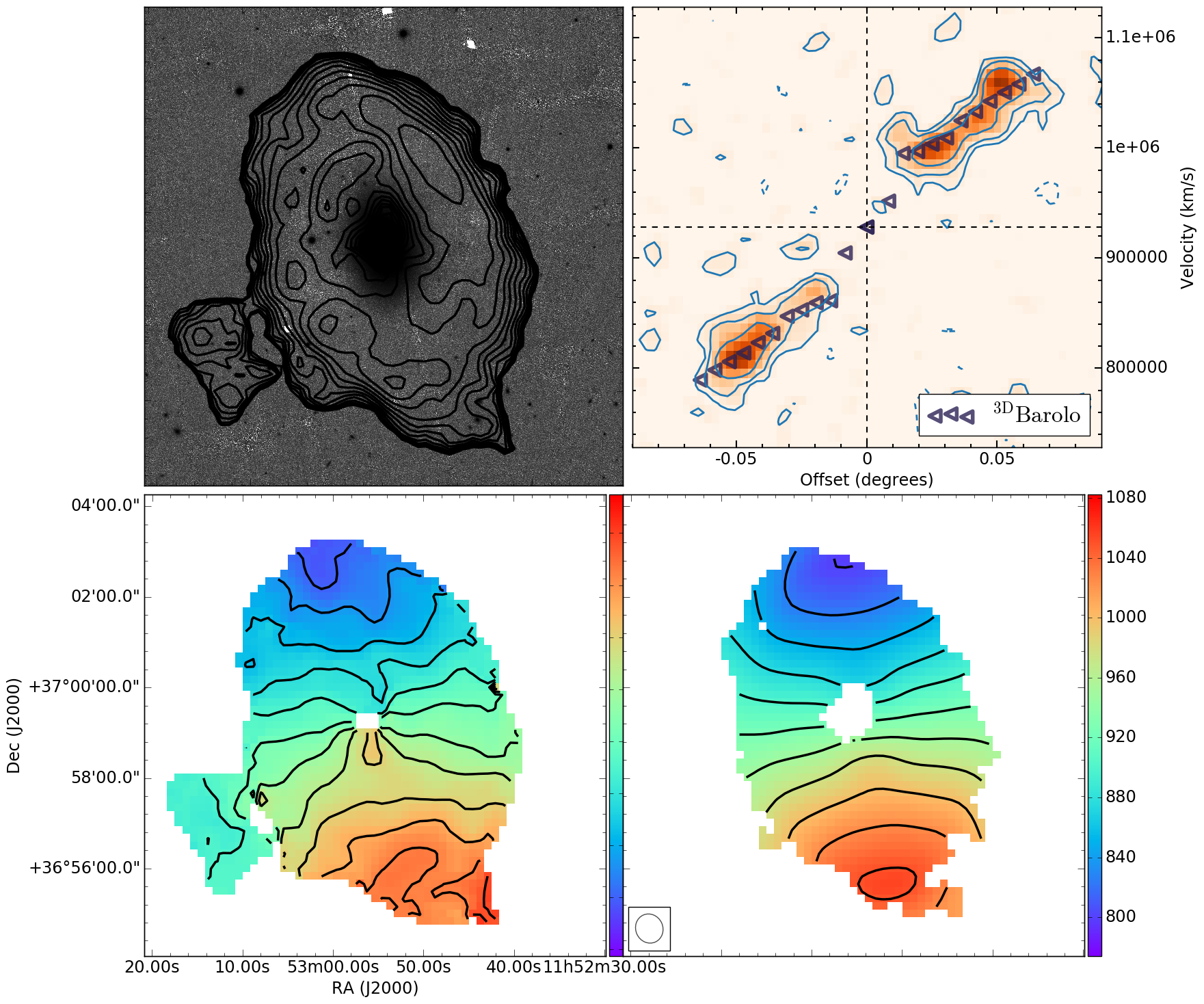}
\caption{H{\sc i} data and kinematics for the galaxy NGC3941. Top left: H{\sc i} distribution of the galaxy overlaid on the SDSS g–band data,Top right: Position velocity diagram (outer contour corresponds to 2$\sigma$) with the rotation curves overlaid on them. Bottom left: velocity field of data, and Bottom right:velocity field of the best fitting \fat model. Velocity contours run from 800 to 1050 km s$^{-1}$ with a spacing of 25 km s$^{-1}$. }
\label{fig:mom8}
\end{figure}

\begin{figure}
\centering
\includegraphics[width=1.0\linewidth]{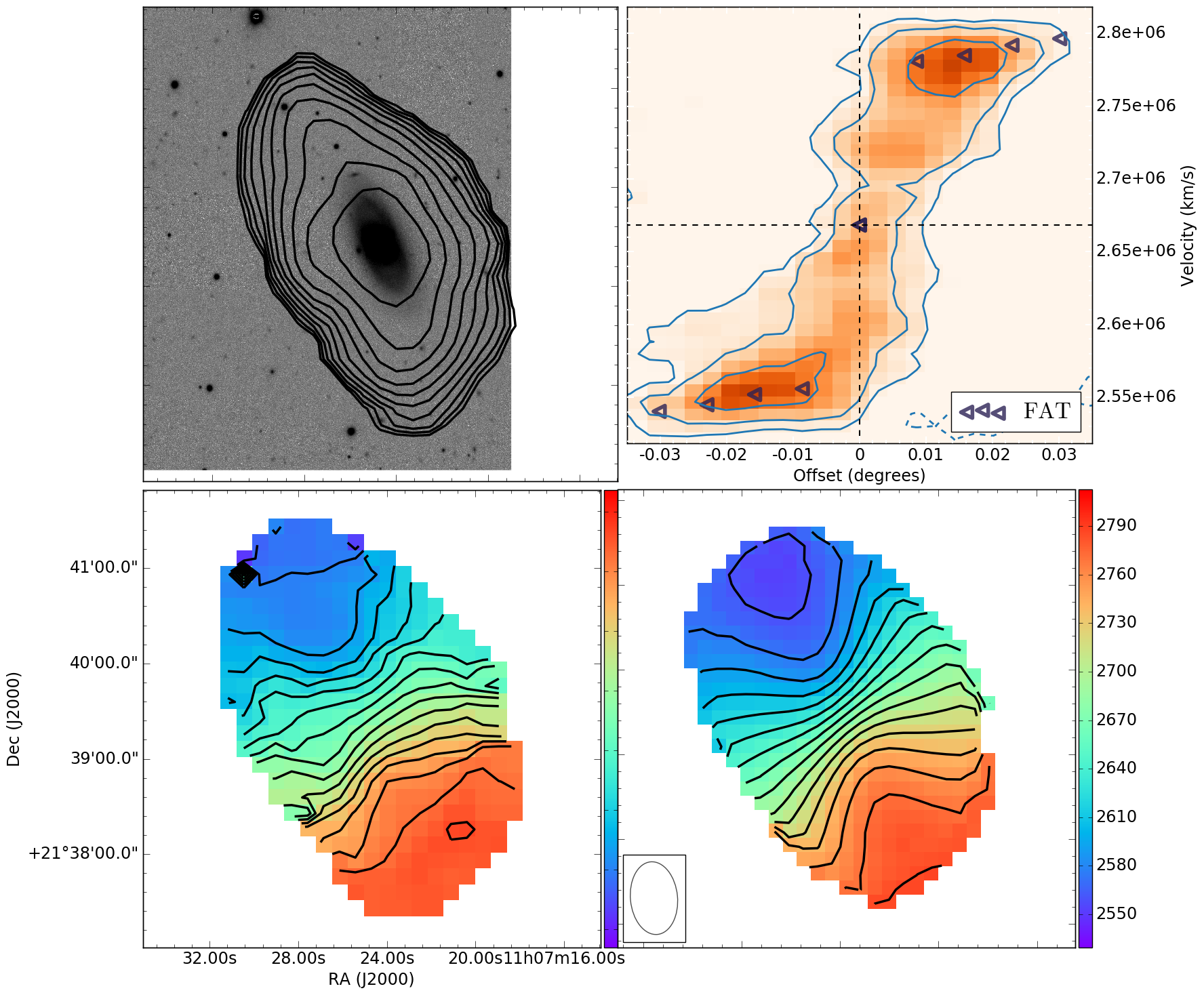}
\caption{H{\sc i} data and kinematics for the galaxy UGC6176. Top left: H{\sc i} distribution of the galaxy overlaid on the SDSS g–band data,Top right: Position velocity diagram (outer contour corresponds to 2$\sigma$) with the rotation curves overlaid on them. Bottom left: velocity field of data, and Bottom right:velocity field of the best fitting \fat model. Velocity contours run from 2560 to 2800 km s$^{-1}$ with a spacing of 15 km s$^{-1}$. }
\label{fig:mom9}
\end{figure}

\begin{figure}
\centering
\includegraphics[width=1.0\linewidth]{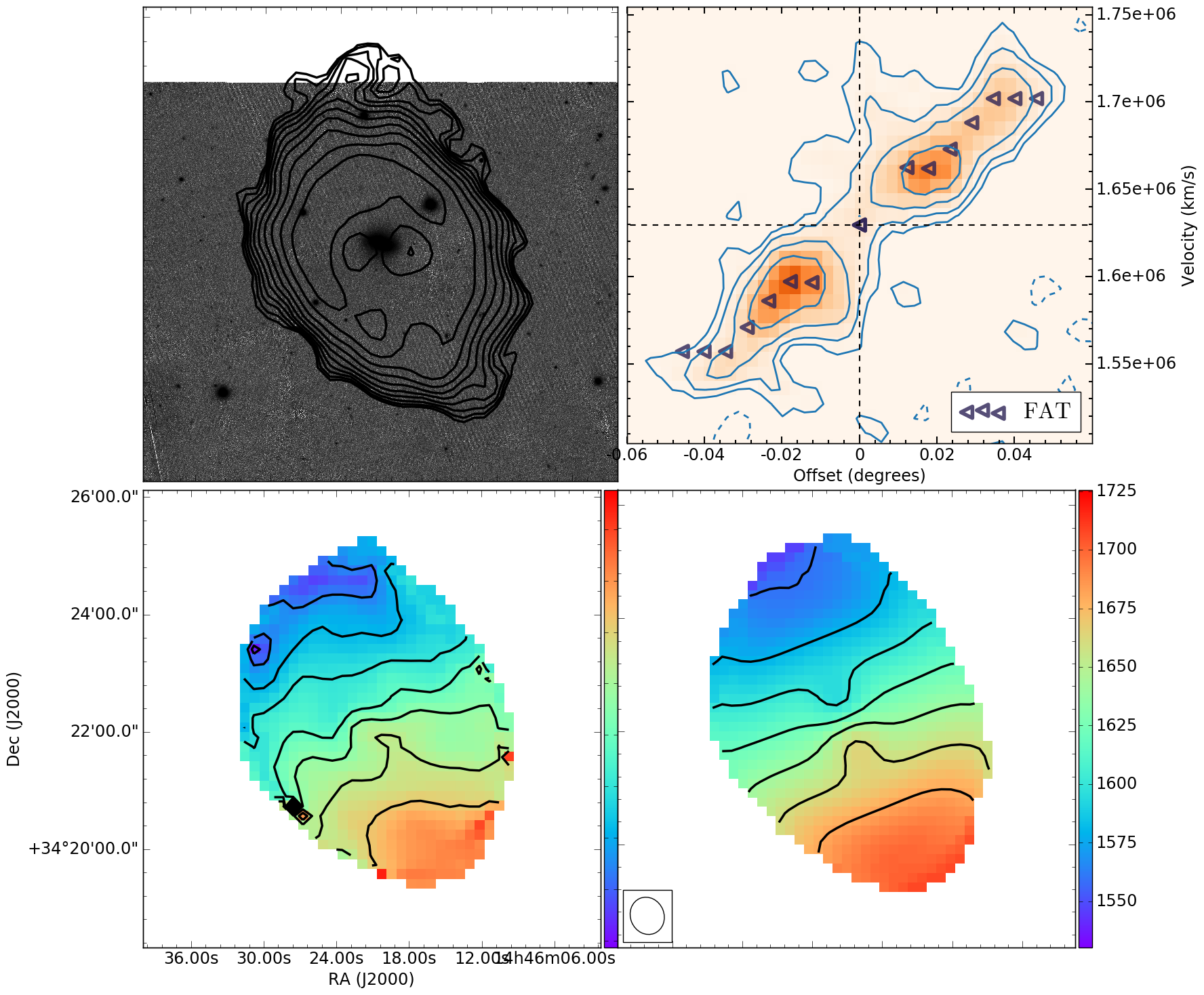}
\caption{H{\sc i} data and kinematics for the galaxy UGC9519. Top left: H{\sc i} distribution of the galaxy overlaid on the SDSS g–band data,Top right: Position velocity diagram (outer contour corresponds to 2$\sigma$) with the rotation curves overlaid on them. Bottom left: velocity field of data, and Bottom right:velocity field of the best fitting \fat model. Velocity contours run from 1560 to 1680 km s$^{-1}$ with a spacing of 20 km s$^{-1}$. }
\label{fig:mom10}
\end{figure}

In this appendix, we present the data and the kinematic analysis of 2 H{\sc i}-- bearing ultra diffuse  galaxies and 8 early--type galaxies. In Fig. \ref{fig:mom1} - \ref{fig:mom10},  we show (i) the integrated H{\sc i} intensity map contours overlaid on the optical image, (ii) Position-velocity diagram taken along the major axis of the galaxy with the rotation curves overlaid on them. The dashed lines indicate the systemic velocity and kinematic center. The overplotted  violet triangles represent the rotation curve derived by 'FAT' (iii) the intensity weighted first moment of the  galaxy,  and  (iv)  velocity  field  of  the  best  fitting FAT model for all the HUDs and early-type galaxies.


\begin{figure}
\centering
\includegraphics[width=0.4\linewidth]{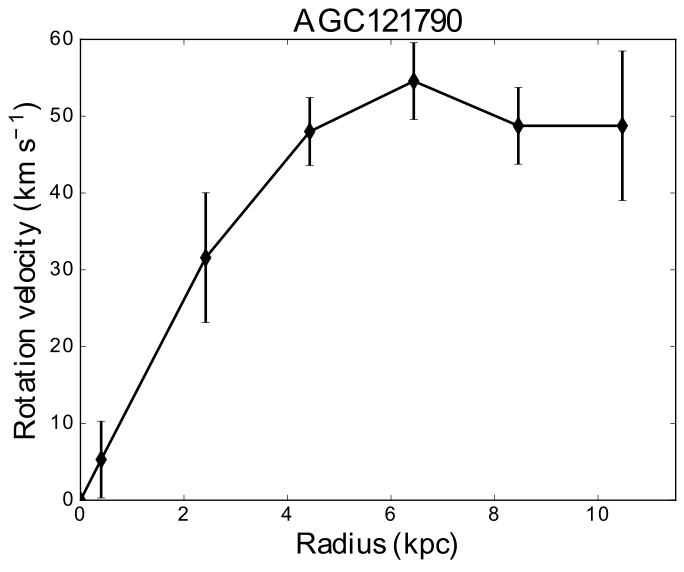}
\includegraphics[width=0.4\linewidth]{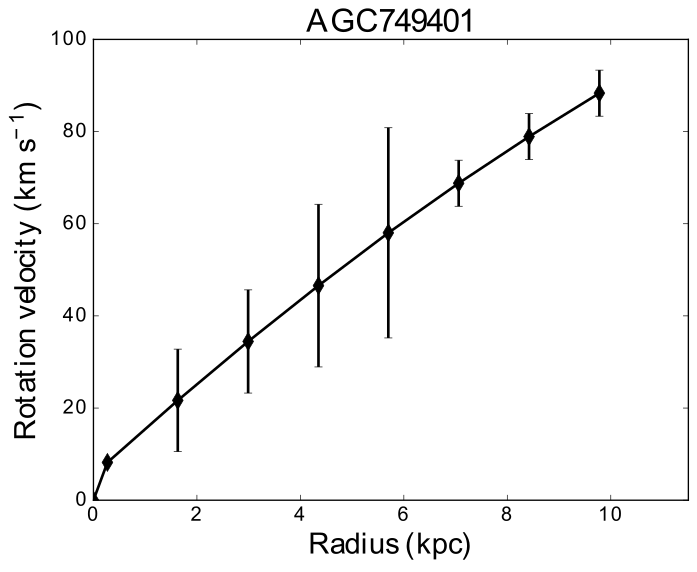}
\caption{Rotation curves of ultra-diffuse galaxies, AGC121790 and AGC749401.}
\label{fig:rotcur1}
\end{figure}

\begin{figure}
\centering
\includegraphics[width=0.48\linewidth]{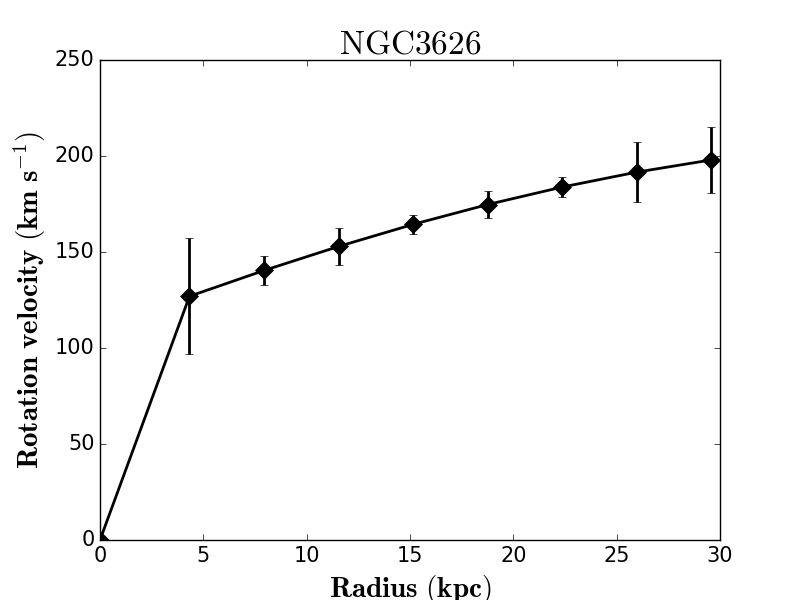}
\includegraphics[width=0.48\linewidth]{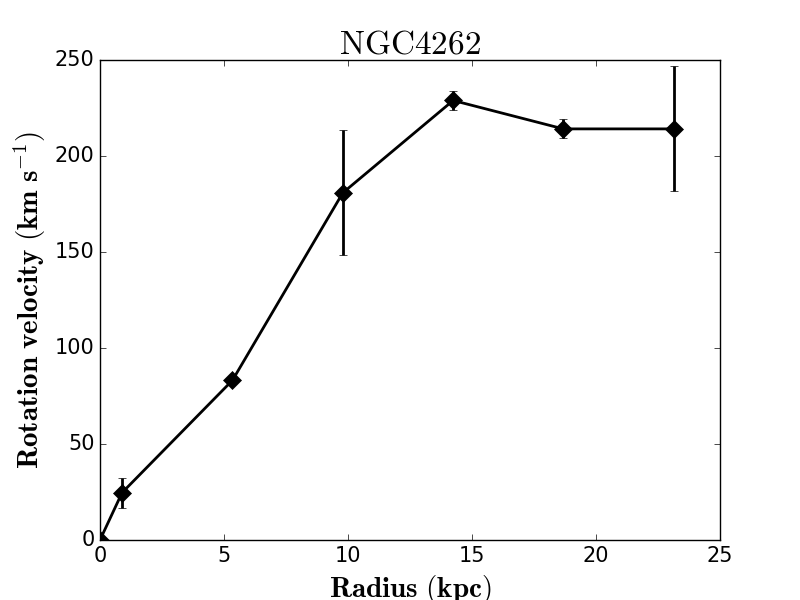}
\includegraphics[width=0.48\linewidth]{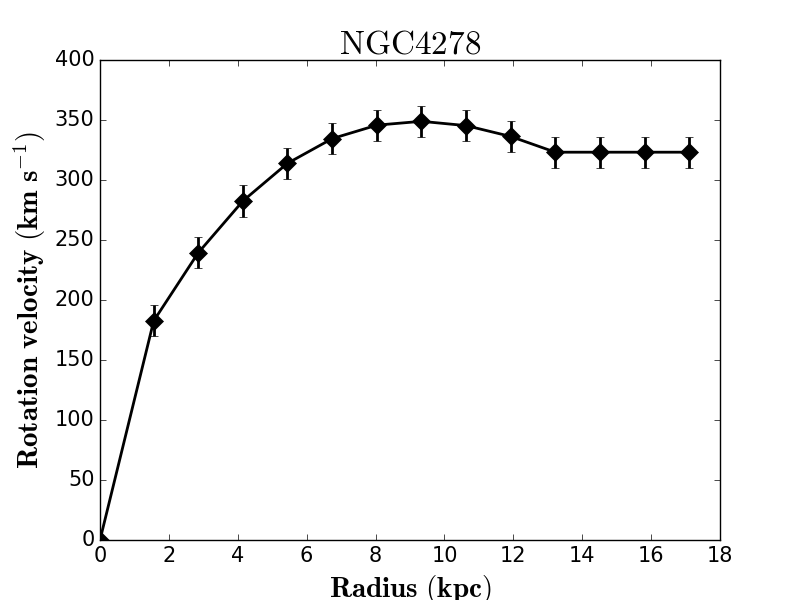}
\includegraphics[width=0.48\linewidth]{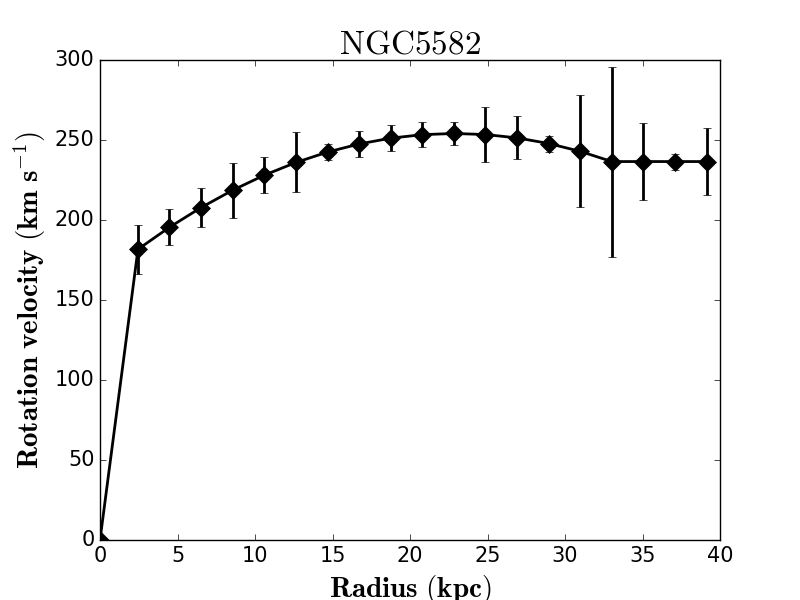}
\includegraphics[width=0.48\linewidth]{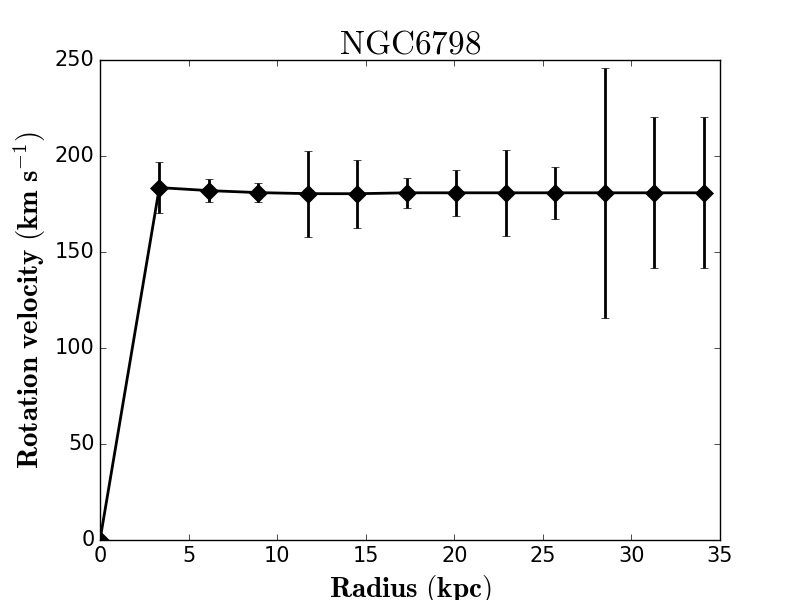}
\includegraphics[width=0.48\linewidth]{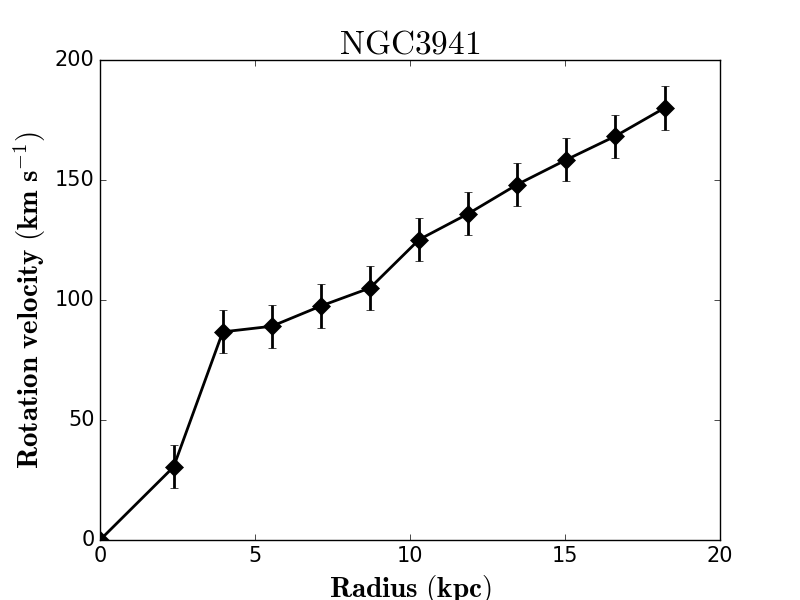}
\includegraphics[width=0.48\linewidth]{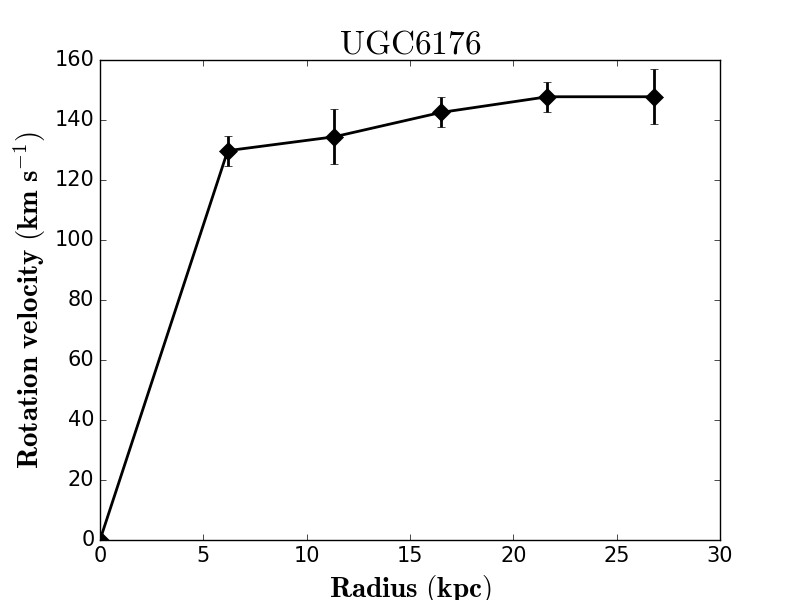}
\includegraphics[width=0.48\linewidth]{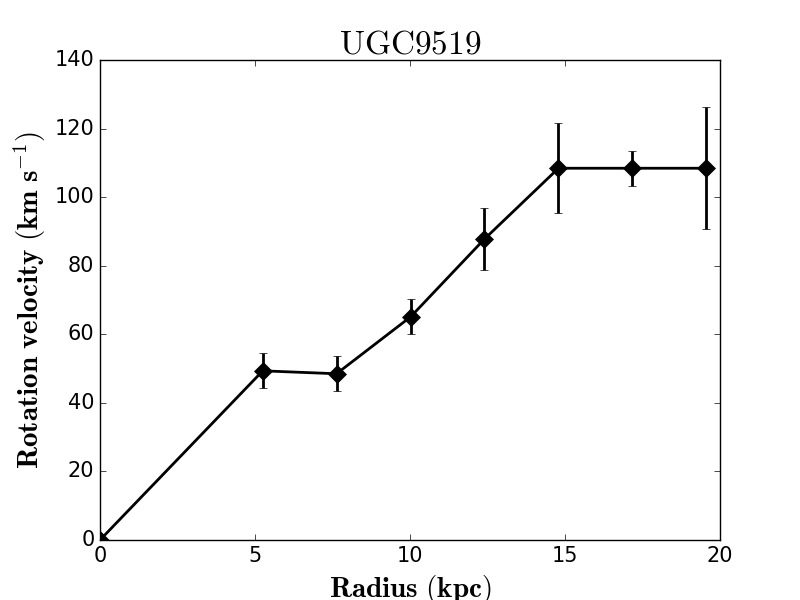}
\caption{Rotation curves of early--type galaxies. Error bars on NGC4278 are taken to be the velocity resolution of the data cube (13 km s$^{-1}$, which is greater than median of FAT errors of all galaxies) since FAT error bars were unreliable.}
\label{fig:rotcur2}
\end{figure}
\begin{table*}
\begin{footnotesize}

\caption{Measured values of specific angular momentum and mass (stars, gas, and baryons) for ultra--diffuse galaxies in this study }
\label{Table1}
\begin{tabular}{ p{1.3cm} p{0.7cm} p{1.8cm}  p{1.8cm} p{1.8cm} p{2.2cm} p{2.2cm}  p{2.2cm}}
\\
\hline
 Galaxy  & dist& M$_{\rm s}$ & M$_{\rm g}$ & M$_{\rm bar}$ & j$_{\rm s}$ & j$_{\rm g}$ & j$_{\rm bar}$\\
         &  (Mpc)  &(log$_{10}$ M$_{\odot}$) & (log$_{10}$ M$_{\odot}$) & (log$_{10}$ M$_{\odot}$) & (log$_{10}$ kpc km/s) & (log$_{10}$ kpc km/s) & (log$_{10}$ kpc km/s) \\
\hline
AGC749401	& 41.8& 7.31	$\pm$ 0.131	& 8.66	$\pm$ 0.137	& 8.68	$\pm$ 0.131 & 2.12	$\pm$ 0.110	& 2.39	$\pm$ 0.093 & 2.39 $\pm$ 0.092 \\[1.0ex]	
AGC121790	& 37.5& 6.40	$\pm$ 0.131	& 8.89	$\pm$ 0.137	& 8.89	$\pm$ 0.136 & 2.00	$\pm$ 0.081	& 2.46	$\pm$ 0.074 & 2.46 $\pm$ 0.075 \\[1.0ex]	
\hline
\end{tabular}
\end{footnotesize}
\end{table*}

\begin{figure}
\centering
\includegraphics[width=0.48\linewidth]{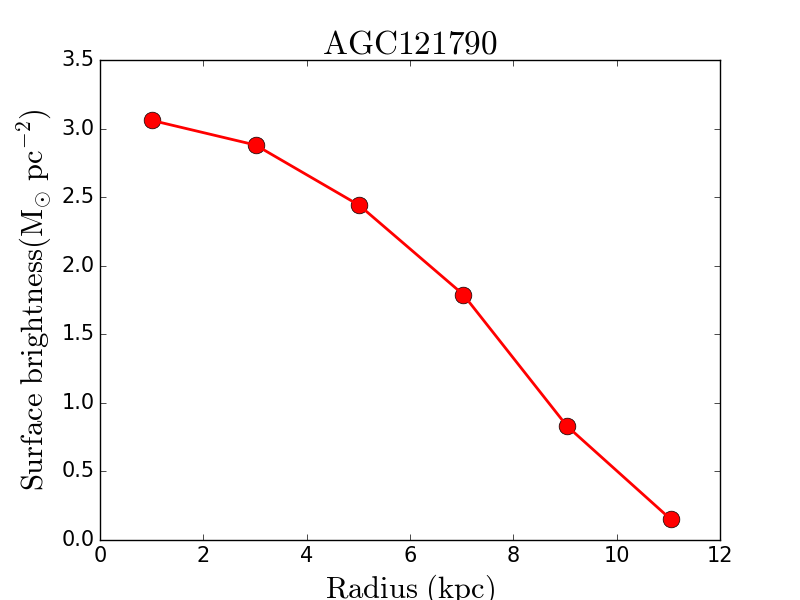}
\includegraphics[width=0.48\linewidth]{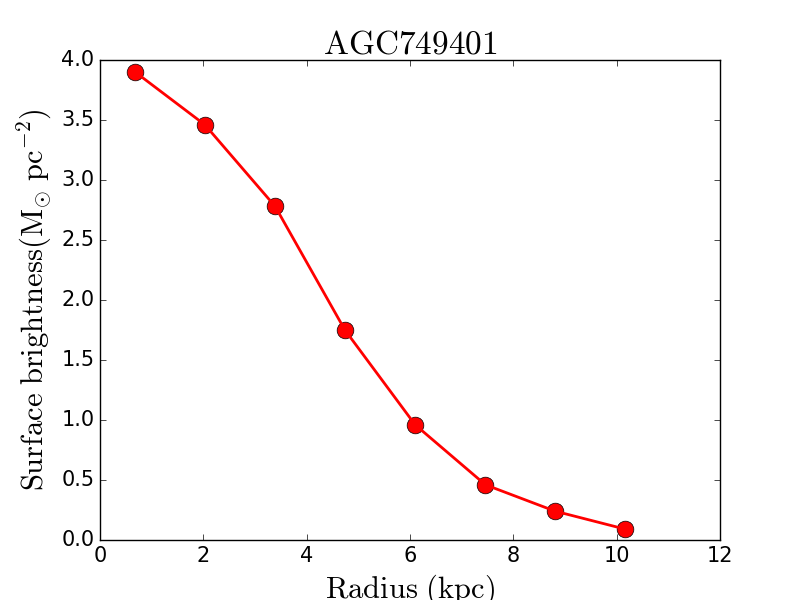}
\caption{H{\sc i} surface brightness as a function of radius for the ultra-diffuse galaxies AGC121790, and AGC749401 respectively.}
\label{fig:sbr1}
\end{figure}

\begin{table*}
\begin{footnotesize}

\caption{Measured values of specific angular momentum and mass (stars, gas, and baryons) for UMa spiral galaxies in this study}
\label{Table2}
\begin{tabular}{ p{1.3cm} p{0.7cm} p{1.8cm}  p{1.8cm} p{1.8cm} p{2.2cm} p{2.2cm}  p{2.2cm}}
\\
\hline
 Galaxy  & dist& M$_{\rm s}$ & M$_{\rm g}$ & M$_{\rm bar}$ & j$_{\rm s}$ & j$_{\rm g}$ & j$_{\rm bar}$\\
         &  (Mpc)  &(log$_{10}$ M$_{\odot}$) & (log$_{10}$ M$_{\odot}$) & (log$_{10}$ M$_{\odot}$) & (log$_{10}$ kpc km/s) & (log$_{10}$ kpc km/s) & (log$_{10}$ kpc km/s) \\
\hline
NGC3877	& 15.27	& 10.6	$\pm$ 0.170	& 9.14	$\pm$ 0.137	& 10.62	$\pm$ 0.164		& 2.87	$\pm$ 0.065		& 3.04	$\pm$ 0.142		& 2.88	$\pm$ 0.066\\[1.0ex]	
NGC3917	& 17.14	& 10.13	$\pm$ 0.170	& 9.34	$\pm$ 0.137	& 10.19	$\pm$ 0.147		& 2.75	$\pm$ 0.065		& 3.07	$\pm$ 0.169		& 2.81	$\pm$ 0.077\\[1.0ex]	
NGC3953	& 18.82	& 10.73	$\pm$ 0.170	& 9.62	$\pm$ 0.137	& 10.76	$\pm$ 0.158		& 3.14	$\pm$ 0.067		& 3.46	$\pm$ 0.120		& 3.17	$\pm$ 0.068\\[1.0ex]	
NGC3972	& 20.04	& 10.10	$\pm$ 0.170	& 9.31	$\pm$ 0.137	& 10.16	$\pm$ 0.147		& 2.70	$\pm$ 0.066		& 2.98	$\pm$ 0.132		& 2.75	$\pm$ 0.071\\[1.0ex]	
NGC3992	& 27.04	& 10.91	$\pm$ 0.170	& 10.23	$\pm$ 0.137	& 11.0	$\pm$ 0.143		& 3.36	$\pm$ 0.068		& 3.84	$\pm$ 0.068		& 3.49	$\pm$ 0.068\\[1.0ex]	
NGC4013	& 19.23	& 10.81	$\pm$ 0.170	& 9.65	$\pm$ 0.137	& 10.84	$\pm$ 0.159		& 2.93	$\pm$ 0.065		& 3.42	$\pm$ 0.065		& 2.99	$\pm$ 0.065\\[1.0ex]	
NGC4100	& 20.04	& 10.62	$\pm$ 0.170	& 9.74	$\pm$ 0.137	& 10.67	$\pm$ 0.151		& 3.01	$\pm$ 0.066		& 3.39	$\pm$ 0.156		& 3.07	$\pm$ 0.074\\[1.0ex]	
NGC4157	& 16.29	& 10.87	$\pm$ 0.170	& 9.96	$\pm$ 0.137	& 10.92	$\pm$ 0.152		& 2.92	$\pm$ 0.065		& 3.40	$\pm$ 0.097		& 3.01	$\pm$ 0.068\\[1.0ex]	
NGC4183	& 17.21	& 10.12	$\pm$ 0.170	& 9.65	$\pm$ 0.137	& 10.25	$\pm$ 0.132		& 2.76	$\pm$ 0.066		& 3.06	$\pm$ 0.137		& 2.85	$\pm$ 0.082\\[1.0ex]	
NGC4217	& 18.19	& 10.75	$\pm$ 0.170	& 9.53	$\pm$ 0.137	& 10.78	$\pm$ 0.161		& 2.97	$\pm$ 0.065		& 3.27	$\pm$ 0.114		& 2.99	$\pm$ 0.066\\[1.0ex]	
UGC6399	& 22.08	& 9.48	$\pm$ 0.171	& 9.19	$\pm$ 0.137	& 9.66	$\pm$ 0.122		& 2.54	$\pm$ 0.069		& 2.79	$\pm$ 0.141		& 2.64	$\pm$ 0.091\\[1.0ex]	
UGC6446	& 18.03	& 8.83	$\pm$ 0.172	& 9.62	$\pm$ 0.137	& 9.69	$\pm$ 0.121		& 2.29	$\pm$ 0.088		& 2.88	$\pm$ 0.100		& 2.83	$\pm$ 0.112\\[1.0ex]	
UGC6667	& 18.45	& 9.80	$\pm$ 0.170	& 9.06	$\pm$ 0.137	& 9.87	$\pm$ 0.145		& 2.67	$\pm$ 0.065		& 2.70	$\pm$ 0.125		& 2.67	$\pm$ 0.068\\[1.0ex]	
UGC6917	& 20.41	& 9.42	$\pm$ 0.171	& 9.54	$\pm$ 0.137	& 9.78	$\pm$ 0.107		& 2.67	$\pm$ 0.073		& 2.99	$\pm$ 0.117		& 2.88	$\pm$ 0.101\\[1.0ex]	
UGC6983	& 21.28	& 9.29	$\pm$ 0.170	& 9.74	$\pm$ 0.137	& 9.87	$\pm$ 0.111		& 2.62	$\pm$ 0.071		& 3.08	$\pm$ 0.095		& 3.00	$\pm$ 0.091\\[1.0ex]	
UGC7089	& 11.50	& 9.06	$\pm$ 0.171	& 8.80	$\pm$ 0.137	& 9.25	$\pm$ 0.120		& 2.19	$\pm$ 0.067		& 2.54	$\pm$ 0.066		& 2.34	$\pm$ 0.067\\[1.0ex]		
\hline

\end{tabular}
\end{footnotesize}
\end{table*}

\begin{table}
\begin{footnotesize}

\caption{Measured values of gas mass and gas specific angular momentum of 6 S0 and 2 elliptical galaxies in this study}
\label{Table3}
\begin{tabular}{ p{1.0cm} p{0.5cm} p{0.5cm} p{1.6cm} p{2.1cm} p{0.7cm} }
\\
\hline
Galaxy & Morp		& dist  & M$_{\rm g}$ &  j$_{\rm g}$ & Notes  \\ 
 & & (Mpc) & (log$_{10}$ M$_{\odot}$) & (log$_{10}$~kpc~km~s$^{-1}$) & \\
\hline
NGC3626	& S0	& 19.5	& 9.19 $\pm$ 0.137	& 3.60	$\pm$ 0.073 & C \\[1.0ex]		
NGC4262	& S0	& 15.4	& 8.84 $\pm$ 0.137	& 3.27	$\pm$ 0.086 & R \\[1.0ex]		
NGC4278	& E	    & 15.6	& 8.90 $\pm$ 0.137	& 3.64	$\pm$ 0.091 & M, L \\[1.0ex]		
NGC5582	& E  	& 27.7	& 9.78 $\pm$ 0.137	& 3.85	$\pm$ 0.074 & R \\[1.0ex]		
NGC3941	& S0	& 11.9	& 8.86 $\pm$ 0.137	& 3.15	$\pm$ 0.073 & C, R \\[1.0ex]		
NGC6798	& S0	& 37.5	& 9.64 $\pm$ 0.137	& 3.67	$\pm$ 0.077 & C  \\[1.0ex]		
UGC6176	& S0	& 40.1	& 9.21 $\pm$ 0.137	& 3.21	$\pm$ 0.068 & W \\[1.0ex]		
UGC9519	& S0	& 27.6	& 9.40 $\pm$ 0.137	& 3.02	$\pm$ 0.081 & M \\[1.0ex]	
\hline
\multicolumn{5}{l}{ Notes on H{\sc i} morphology and kinematics:}\\
\multicolumn{5}{l}{ C = H{\sc i} counter-rotating relative to the stellar kinematics;  } \\
\multicolumn{5}{l}{L= lopsided H{\sc i} morphology; M= H{\sc i} misaligned relative to } \\ 
\multicolumn{5}{l}{ the stellar kinematics; R= ring; W= warp}\\

\end{tabular}
\end{footnotesize}
\end{table}  

\begin{table}
\begin{footnotesize}

\caption{Convergence factors of 16 UMa spirals for \jb and \jg profiles. }
\label{Table4}
\begin{tabular}{ p{1.5cm} p{1.5cm} p{1.5cm} }
\\
\hline
Galaxy & 	r$_{\rm bar}$	& r$_{\rm g}$  \\ 
\hline
N3877   & 0.90 & 0.90 \\[1.0ex]
N3917   & 0.94 & 0.86 \\[1.0ex]
N3953   & 0.98 & 0.90 \\[1.0ex]
N3972   & 0.94 & 0.82 \\[1.0ex]
N3992   & 0.97 & 0.95 \\[1.0ex]
N4013   & 0.97 & 0.88 \\[1.0ex]
N4100   & 0.96 & 0.90 \\[1.0ex]
N4157   & 0.95 & 0.88 \\[1.0ex]
N4183   & 0.91 & 0.82 \\[1.0ex]
N4217   & 0.97 & 0.83 \\[1.0ex] 
U6399   & 0.86 & 0.78 \\[1.0ex]
U6446   & 0.94 & 0.94 \\[1.0ex]
U6667   & 0.66 & 0.73 \\[1.0ex]
U6917   & 0.89 & 0.88 \\[1.0ex]
U6983   & 0.94 & 0.94 \\[1.0ex]
U7089   & 0.82 & 0.86 \\[1.0ex]
\hline
\end{tabular}
\end{footnotesize}
\end{table} 

\begin{table}
\begin{footnotesize}
\caption{Convergence factors of 2 ultra-diffuse galaxies for \jb and \jg profiles.}
\label{Table5}
\begin{tabular}{ p{1.5cm} p{1.5cm} p{1.5cm} }
\\
\hline
Galaxy & 	r$_{\rm bar}$	& r$_{\rm g}$  \\ 
\hline
AGC121790   & 0.74 & 0.74 \\[1.0ex]
AGC749401   & 0.61 & 0.60 \\[1.0ex]
\hline
\end{tabular}
\end{footnotesize}
\end{table}  

\begin{table}
\begin{footnotesize}
\caption{Convergence factors of 8 early-type galaxies for \jg profiles. }
\label{Table6}
\begin{tabular}{p{3.5cm} p{2.5cm}}
\hline
Galaxy &  r$_{\rm g}$  \\ 
\hline
NGC4262 & 0.80 \\
NGC4278 & 0.71 \\
NGC3626 & 0.61 \\
NGC3941 & 0.89 \\
NGC5582 & 0.85 \\
NGC6798 & 0.43 \\
UGC9519 & 0.74 \\
UGC6176 & 0.85 \\
\hline
\end{tabular}
\end{footnotesize}
\end{table} 

\begin{figure}
\centering
\includegraphics[width=0.48\linewidth]{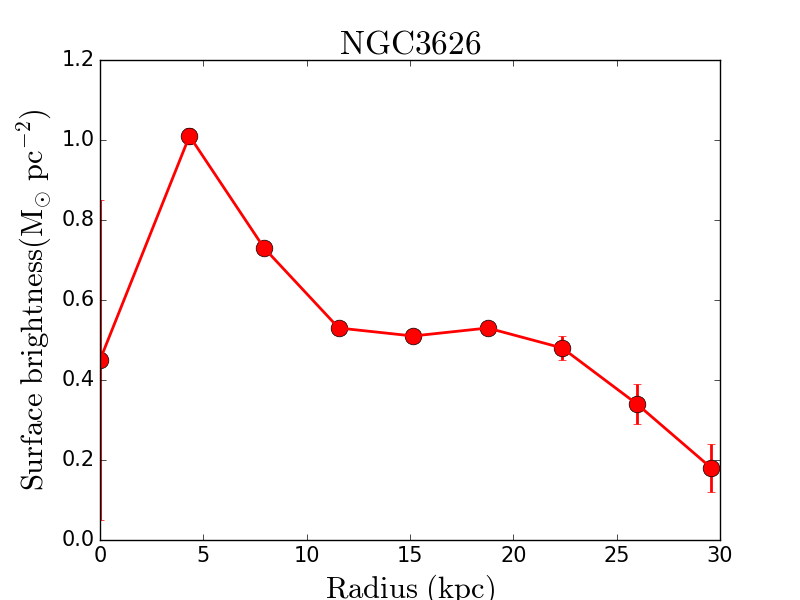}
\includegraphics[width=0.48\linewidth]{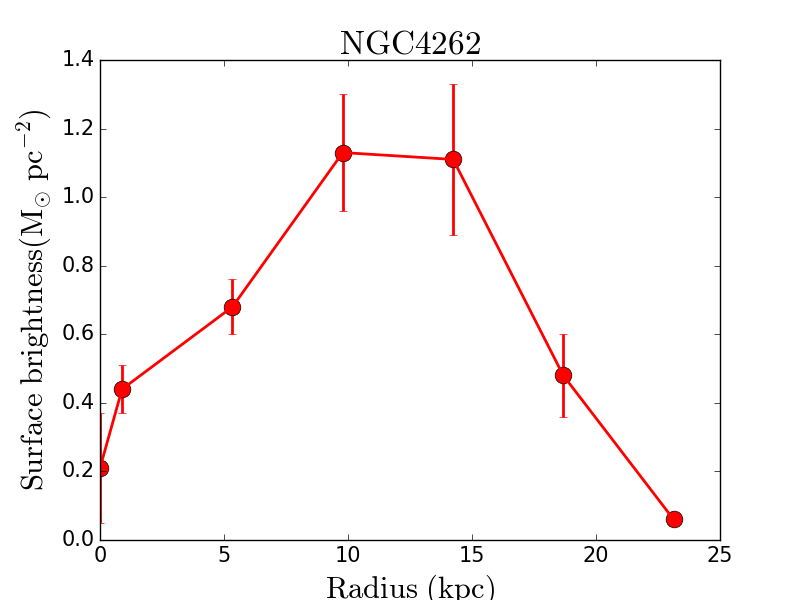}
\includegraphics[width=0.48\linewidth]{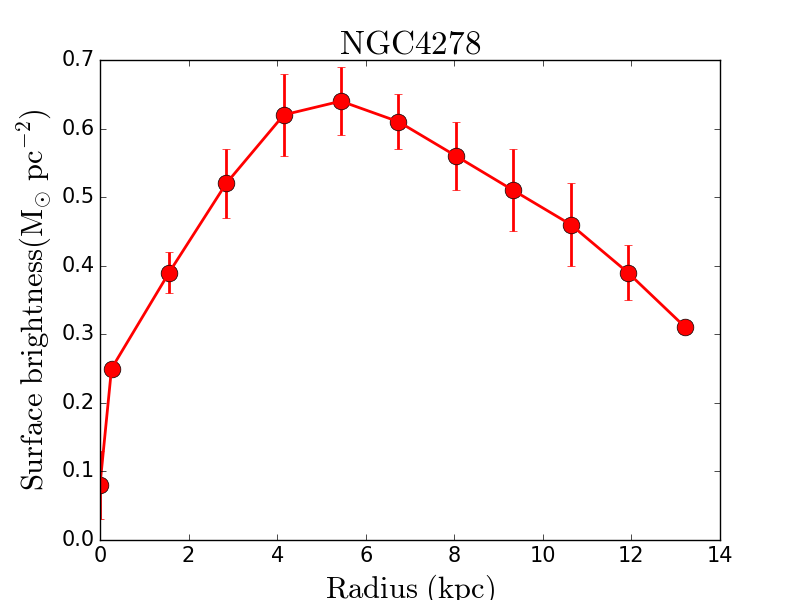}
\includegraphics[width=0.48\linewidth]{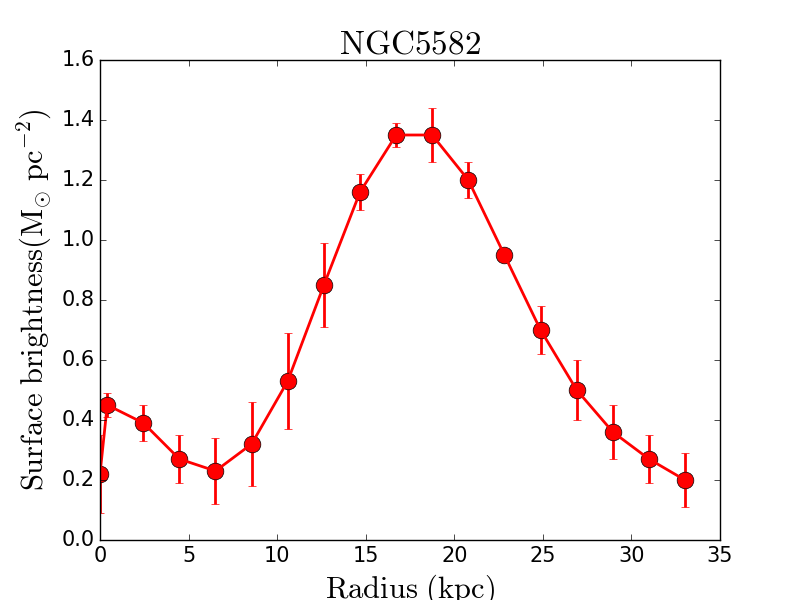}
\includegraphics[width=0.48\linewidth]{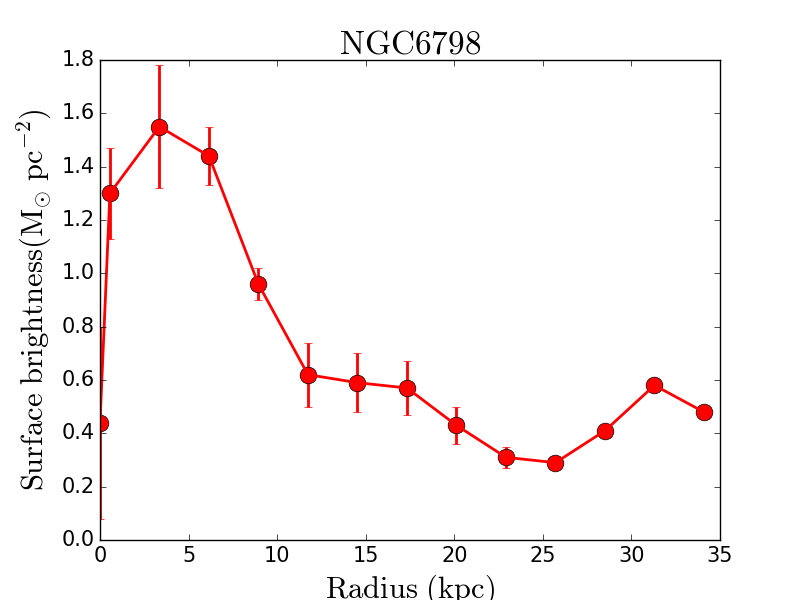}
\includegraphics[width=0.48\linewidth]{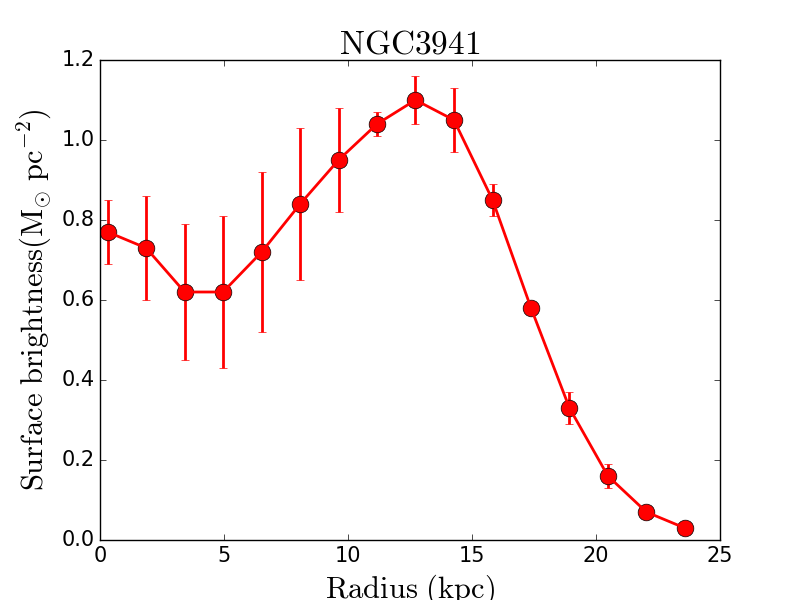}
\includegraphics[width=0.48\linewidth]{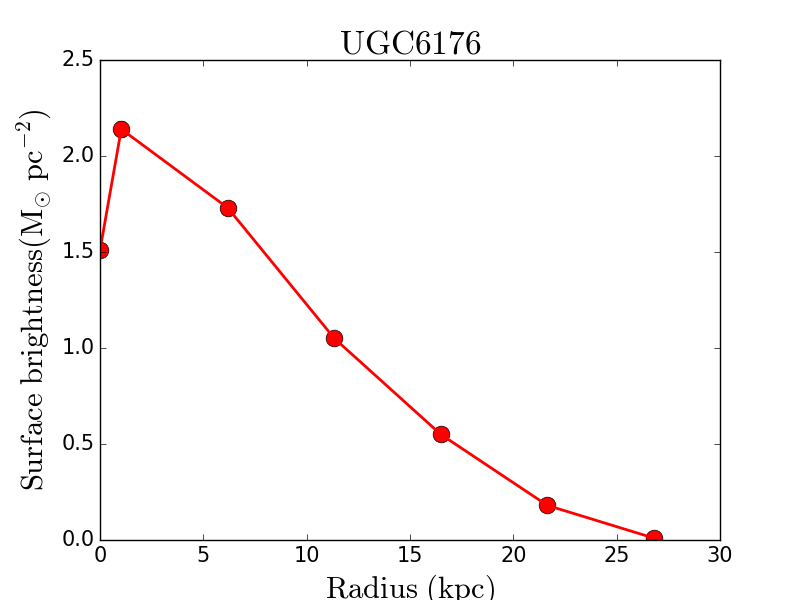}
\includegraphics[width=0.48\linewidth]{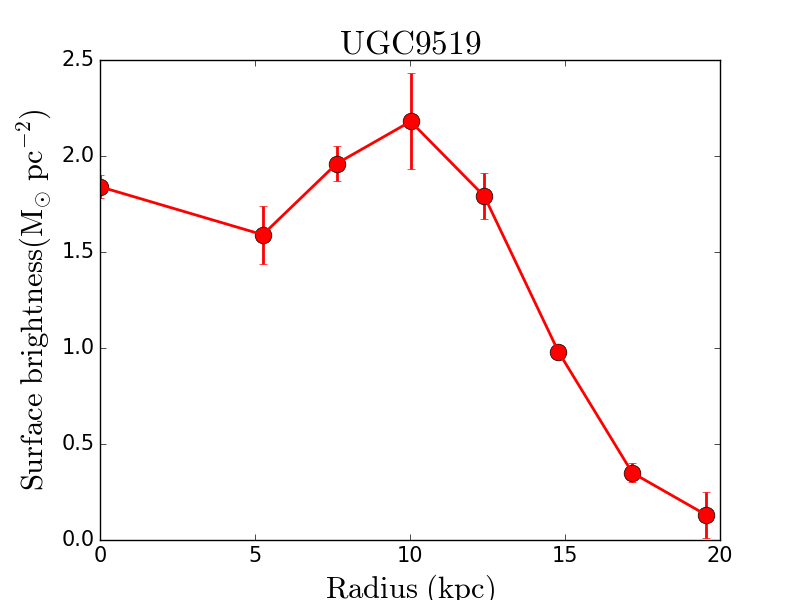}
\caption{H{\sc i} surface brightness as a function of radius for the early-type galaxies NGC3626, NGC4262, NGC4278, NGC5582, NGC6798, NGC3941, UGC6176, and UGC9519 respectively.}
\label{fig:sbr2}
\end{figure}

\begin{figure}
\centering
\includegraphics[width=1.0\linewidth]{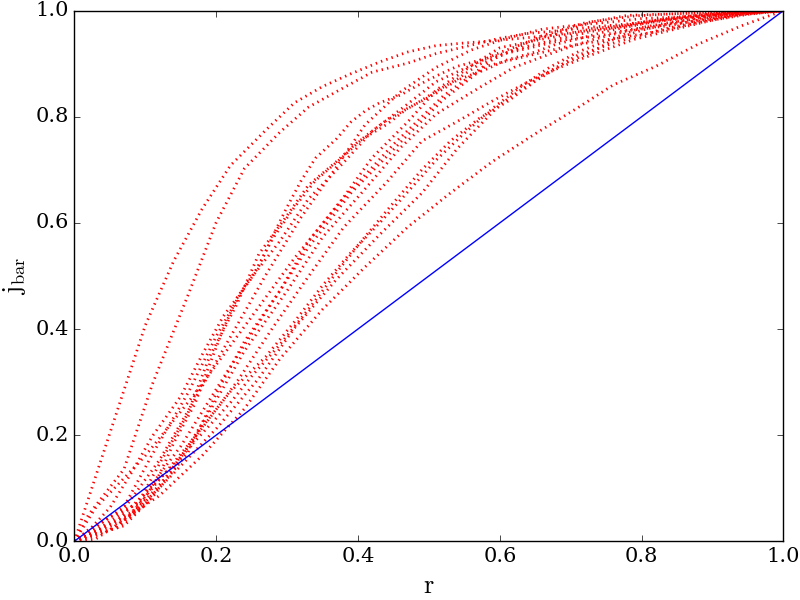}
\includegraphics[width=1.0\linewidth]{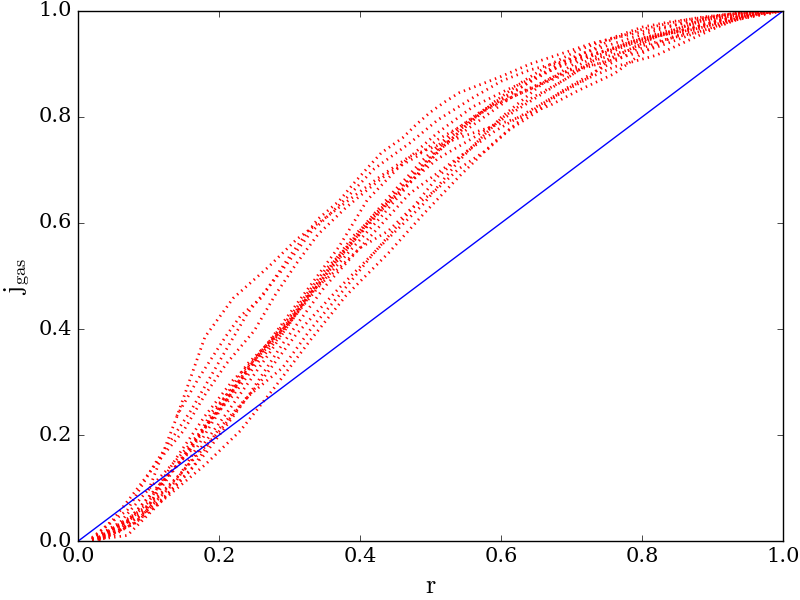}
\caption{The cumulative j$_{\rm bar}$ and j$_{\rm g}$ profiles of 16 UMa spirals. We normalize the axes to allow the comparison between the profiles. The one to one relation is shown with a blue solid line.}
\label{fig:prof1}
\end{figure}

\begin{figure}
\centering
\includegraphics[width=1.0\linewidth]{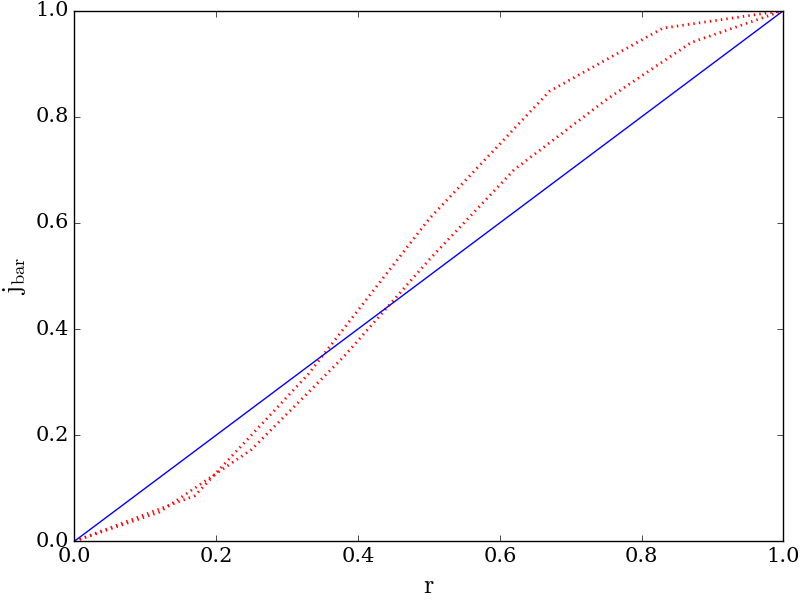}
\includegraphics[width=1.0\linewidth]{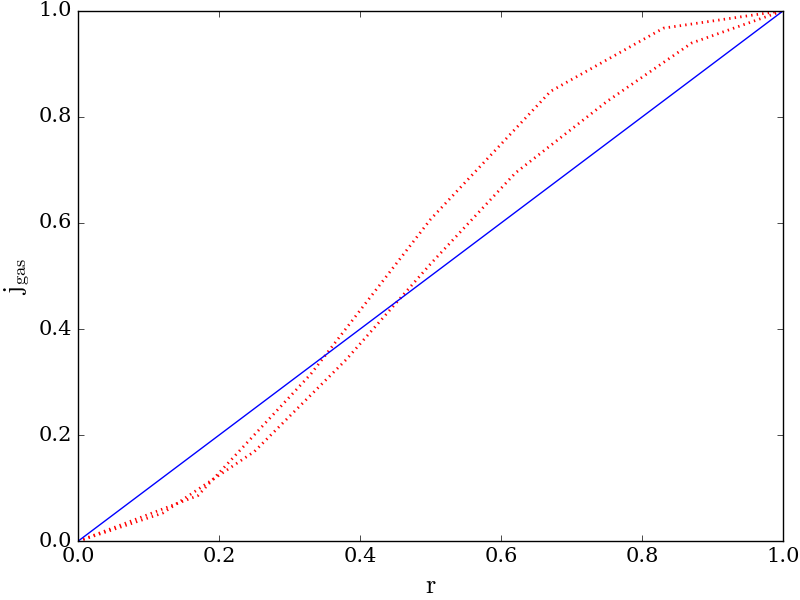} 
\caption{ The cumulative j$_{\rm bar}$ and j$_{\rm g}$ profiles of 2 ultra-diffuse galaxies. We normalize the axes to allow the comparison between the profiles. The one to one relation is shown with a blue solid line.}
\label{fig:prof2}
\end{figure}

\begin{figure}
\centering
\includegraphics[width=1.0\linewidth]{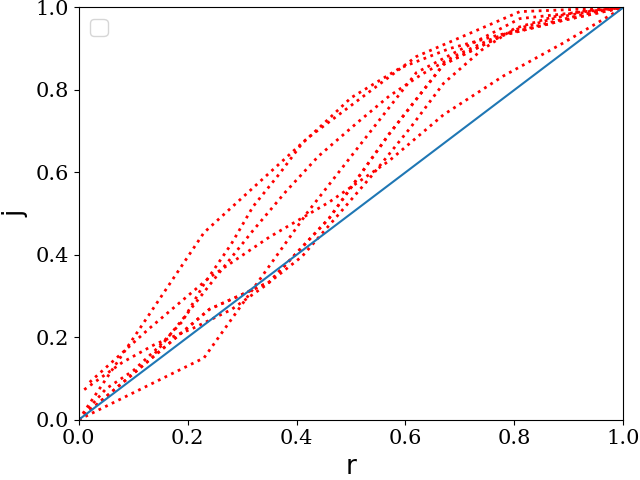}
\caption{The cumulative j$_{\rm g}$ profiles of of 8 early-type galaxies. We normalize the axes to allow the comparison between the profiles. The one to one relation is shown with a blue solid line.}
\label{fig:prof3}
\end{figure}




\bsp	
\label{lastpage}
\end{document}